\documentstyle[12pt]{article}

\def\df #1. #2\par{\medbreak
  \noindent{{\tt {\bf Definition #1.}}\enspace}{\sl#2\par}%
  \ifdim\lastskip<\medskipamount \removelastskip\penalty55\medskip\fi}

\def\theorem #1. #2\par{\medbreak
  \noindent{\tt {\bf Theorem #1.}\enspace}{\sl#2\par}%
  \ifdim\lastskip<\medskipamount \removelastskip\penalty55\medskip\fi}

\def\lemma #1. #2\par{\medbreak
  \noindent{\tt {\bf Lemma #1.}\enspace}{\sl#2\par}%
  \ifdim\lastskip<\medskipamount \removelastskip\penalty55\medskip\fi}

\def\example{\medbreak\noindent{\bf Example}}

\def\proof{\medbreak\noindent{\bf Proof}}

\def\remark{\medbreak\noindent{\bf Note}}

\def\dsl{\raise.15ex\hbox{/}\kern-.65em\nabla}

\def\cl{{\cal C}\!\ell}

\def\lra{\leftrightarrow}
\def\fin{\hbox{$\bullet$}\medskip}
\def\E{{\cal E}}
\def\L{{\cal L}}
\def\V{{\cal V}}
\def\K{{\cal K}}
\def\R{{\cal R}}
\def\C{{\cal C}}
\def\S{{\cal S}}
\def\t{\tilde}
\def\be{\begin{equation}}
\def\ee{\end{equation}}
\def\LE{\Lambda(\E)}
\def\LkE{\Lambda^k(\E)}

\def\LV{\Lambda(\V)}
\def\LkV{\Lambda^k(\V)}

\def\LCV{\Lambda_{\C}(\V)}
\def\tr{{\rm Tr}}
\def\ww{\wedge\ldots\wedge}
\def\dim{{\rm dim}}
\def\sgn{{\rm sgn}}
\def\det{{\rm det}}
\def\even{{\rm even}}
\def\odd{{\rm odd}}

\def\kfac{\frac{1}{k!}}
\def\spin{{\top_1\Lambda^2_\R}}
\def\spinup{{\top^1\Lambda^2_\R}}
\def\spintwo{{\top_2\Lambda^2_\R}}
\def\uone{{\top_1\Lambda^0_{i\R}}}
\def\uoneup{{\top^1\Lambda^0_{i\R}}}
\def\uonetwo{{\top_2\Lambda^0_{i\R}}}

\def\Spin{{\rm Spin}}
\def\com{{\rm com}}
\def\exp{{\rm exp}}
\def\diag{{\rm diag}}
\def\s{\stackrel}
\def\UV{{\rm U(1)_{\V}}}

\begin{document}

\title{A gauge model with spinor group
for a description of local interaction
of a fermion with electromagnetic and gravitational fields}

\author{N.G.Marchuk  \thanks{Research supported by the Russian Foundation
for Basic Research, grant 00-01-00224}}


\maketitle

Steklov Mathematical Institute,
Gubkina st.8,
Moscow 117966, Russia;
nikolai@marchuk.mian.su; nmarchuk@mi.ras.ru;
http://www.orc.ru/\~{}nmarchuk
\vskip 1cm

\begin{abstract}
We suggest model equations, which, from some point of view,
describe local interaction of three physical fields: a field of
matter, an electromagnetic field and a gravitational field. A base of
the model is a field of matter described by the wave function
of fermion satisfying the equation similar to Dirac equation for
electron. Electromagnetic and gravitational fields appear as the
gauge fields for this equation. We have found the connection between
these fields and the curvature tensor of Riemannian manifold.
We present a main
Lagrangian from which the equations of the model are
deduced. The covariance of the model equations under changes of
coordinates is considered. We develop
mathematical techniques needed for the model
connected with an exterior algebra of Euclidean
or Riemannian space. The exterior algebra is considered as a bialgebra
with two operations of multiplications -- an exterior multiplication and
Clifford multiplication. We define a structure of Euclidean or
Riemannian space on the exterior algebra, which leads to the notions
of Spin-isometric change of coordinates and Spin-isometric manifold used
in the model.

In the revised paper we correct an error with the formula $G_{ij}=-U^{-1}D_{ij}U/2$,
(now $U=1$).
\end{abstract}


\tableofcontents


\section*{Introduction}
The aim of this paper is to introduce a mathematical model, which describes
a local interaction of three physical fields: a field of matter, an
electromagnetic field and a gravitational field. We begin with equations
that were considered in
\cite{marchuk2},\cite{marchuk3},\cite{marchuk},\cite{marchuk1}
for the case of Minkowski space. The main
equation, which is similar to Dirac equation for an electron, considered on
Riemannian manifold $\V$ for the wave function $\Psi$ of a fermion (a field of
matter). The suggested equation is gauge invariant with respect to two groups
-- $U(1)_\V$ and
$\Spin(\V)$. A gauge field which corresponds to  $U(1)_\V$ gauge symmetry is
interpreted as the electromagnetic field. For the gauge field that
corresponds to
$\Spin(\V)$ symmetry  the connection with the curvature tensor of
Riemannian manifold $\V$ is found. This connection gives us the ground to interpret this
gauge field as the gravitational field of physical space. We present the
Lagrangian from which all equations of the model are deduced. The questions
about conservative laws and a covariance of equations with respect to changes of
coordinates are considered. It is shown, that under so called Spin-isometric
changes of coordinates the wave function of fermion $\Psi$ transforms as spinor,
which is needed for the description of fermions. The main equation of the model
was compared with standard Dirac equation for the case of Minkowski space.
\medskip

One can connect the present model with two directions of mathematical physics.
The initiator of the first direction was H.Weyl, who suggested in the year 1918
the model
\cite{weil} of unification of Einstein's theory of general relativity and the
theory of electromagnetic field. Developing this model, H.Weyl
\cite{weil1}, V.A.Fock
\cite{fock} and others had come to the description of electromagnetic field as a
gauge field with
${\rm U(1)}$ Lie group symmetry. The Weyl-Fock principle of gauge symmetry
was generalized in the year 1954 by C.N.Yang and R.L.Mills
\cite{yang} on nonabelian gauge groups and now it is one of basic principles of
theoretical physics. This principle also plays the key role in our model, where
the electromagnetic field and gravitational field appear as gauge fields
corresponding to the groups
$\UV$ and $\Spin(\V)$.
Several authors suggested models which describe the gravity field as a
gauge field with gauge groups like Poincare group, De Sitter group and
so on. The first who developed this gauge approach in the theory of
gravity was, probably, R.~Utiyama \cite{utiyama}. Review see in
\cite{einstein}, \cite{ponomarev}.

After H.Weyl's paper \cite{weil} several authors suggested models of
unification of electromagnetism and gravity. In many of these models
they used different generalizations of Riemannian space: spaces with
affine connectedness, spaces with torsion, spaces with nonsymmetric
metric tensor and so on. We can't review here these models (see, for
example, \cite{einstein},\cite{hehl}).
Our approach to invention of the model, from some point of view, is
opposite. We refuse to use the arbitrary changes of coordinates on
Riemannian manifold and allow to use only Spin-isometric changes of
coordinates. By this we come to Spin-isometric manifold, which is
partial case of Riemannian manifold. Let us note, that suggested model
is essentially different from other known models.
\medskip

The second direction of mathematical physics, with which our model can
be connected, was initiated by P.~A.~M.~Dirac in 1928, when he had invented
a relativistic equation for electron
\cite{dirac}. Several authors suggested generalizations of Dirac's
equation for the case of Riemannian manifold (emerging with presence of
gravity according to Einstein's theory). The main difficulty of such
generalization arises from the fact, that a wave function of electron,
which satisfy Dirac equation, is a spinor, but not a tensor.
Many authors considering such a generalizations of Dirac equation use a
tetrad formalism on
Riemannian manifold \cite{tetrode},\cite{fock2},\cite{hehl}.
But there is another approach, suggested in 1928 by
D.~Ivanenko and L.~Landau \cite{ivanenko} and rediscovered in 1962 by
E.~K\"ahler \cite{kahler}. They had suggested to use, as an equation for
relativistic electron (fermion) on Riemannian manifold, the so called
Ivanenko-Landau-K\"ahler (ILK) equation \footnote{Some authors called it
Dirac-K\"ahler equation}
$$
i(d-\delta)\Psi-m\Psi=0,
$$
where $\Psi$ is a nongomogeneous covariant antisymmetric tensor field
(which is a sum of differential forms of the ranks from $0$ to $4$);
$d$ is the exterior differential and $\delta$ is the operator of
generalized divergence
(chapter 2.6). Similar equations were investigated by A.~A.~Dezin
\cite{dezin}. ILK-equation has an important advantage -- it is covariant
under arbitrary smooth changes of coordinates. But also, it has the
disadvantage -- under a change of coordinates the wave function $\Psi$
transforms as a tensor, but not as a spinor.

The main equation of our model differ from ILK-equation by
terms the presence of which guarantees the gauge invariance of the equation with
respect to the group $\Spin(\V)$.
This new gauge symmetry of the equation is very important, because if we
do a Spin-isometric change of coordinates and accompany it by the
corresponding gauge transformation from the group
$\Spin(\V)$, then, as a result, the wave function $\Psi$
transforms as a spinor, but not as a tensor (chapter 3.4).
\medskip

The paper consists of three parts. A third part is devoted to a
description and investigation of the model. In the first two parts we
collect and systematize the mathematical results needed for the model.

In the first part a main object of interest is a space of exterior forms
$\Lambda(\E)$ of the Euclidean space $\E$. For the forms from
$\LE$, besides the usual operation of exterior multiplication, the new
operation of Clifford multiplication is defined. These two operations of
multiplication induce two basises of linear space
$\LE$ -- Grassmann basis and Clifford basis. The structure of Euclidean
space can be defined on
$\LE$ which allow to define Spin-isometric changes of coordinates on
initial Euclidean space $\E$.

The main construction of exterior algebra $\LE$ with two operations of
multiplication was considered by several authors. The first was
H.~Grassmann \cite{grassmann} in the year 1877.
\medskip

In the second part, the algebraic techniques, developed in the first
part for Euclidean space $\E$, is transferred on elementary Riemannian
manifold $\V$. In addition an analytical aspect of the theory is
considered -- the operators of covariant differentiation
$\nabla_j$ and Clifford differentiation $\Upsilon_k$
are introduced and they are used for the definition of operators
$d,\delta,\Upsilon,\Delta$.

A new notion of Spin-isometric change of coordinates on Riemannian
manifold is invented (chapter 2.4). This notion appear to be important
for the model and it leads to the  notion of Spin-isometric manifold
(chapter 3.4).

The form of presentation of mathematical results in the first and second
parts of paper is essentially different from conventional (for example,
in \cite{benn}).
\medskip

We use the symbol $\bullet$ to mark the end of proof of a theorem, or to
emphasize the absence of proof, that means, that the theorem is proved
by the direct calculation. In formulations of some theorems there are
restrictions of a dimension of space, like $n\leq4$, which indicate that
we can calculate the result only for these $n$. We should like to find
proofs of these theorems which are correct for all natural $n$.

\vfill\eject

\section{An exterior algebra  of Euclidean space}
In the first part of paper we consider some, needed for the model,
mathematical structures, which in the second part transfered to
Riemannian manifold.


\par
\subsection{Euclidean space $\E$.}\par
An $n$-dimensional Euclidean space $\E$ is a complete set of a real
vector space ${\cal R}^n$ and a metric tensor of the second rank $g$.
Covariant componenets of
metric tensor in coordinates  $x^1,\ldots,x^n$ with basis coordinate
vectors  $e_1,\ldots,e_n$ satisfies the following conditions:
\par
\medskip
1. $g_{ij}=g_{ji},\quad i,j=1,\ldots,n$.

2. The matrix $\|g_{ij}\|$ is nondegenerate.
\medskip
\par
In Euclidean space there is a scalar multiplication of vectors
such, that  $(e_i,e_j)=g_{ij}$. Contravariant components of the metric tensor
form a matrix $\|g^{ij}\|$ which is inverse to the matrix $\|g_{ij}\|$.
We do not suppose that matrix $\|g_{ij}\|$ is positive defined.
Basis covectors $e^i=g^{ij}e_j$, $i=1,\ldots,n$ (a sum over $j$ from $1$
to $n$) correspond to the basis vectors $e_1,\ldots,e_n$ and
$(e^i,e^j)=g^{ij}$.
\par
One can consider a linear change of coordinates $(x)\to(\tilde x)$
\begin{equation}
{\tilde x}^i=p^i_j x^j,\quad x^i=q^i_j \tilde x^j,\quad i=1,\ldots,n
\label{D1}
\end{equation}
$$
\frac{\partial x^i}{\partial\t x^j}=q^i_j,\quad
\frac{\partial\t x^i}{\partial x^j}=p^i_j,\quad
p^i_j q^j_k=\delta^i_k,\quad q^i_j p^j_k=\delta^i_k,
$$
where $\delta^i_k$ is Kronecker tensor which is equal to zero, when
$i\neq k$, and equal to $1$, when $i=k$.
In accordance with the rules of a tensor analysis, coordinate vectors
$e_i$, covectors $e^i$ and components of the metric tensor $g$ transform
as
\begin{equation}
\t e_i=q^j_i e_j,\quad \t e^i=p^i_j e^j,\quad
\t g_{i_1 i_2}=q^{j_1}_{i_1}q^{j_2}_{i_2} g_{j_1 j_2},\quad
\t g^{i_1 i_2}=p^{i_1}_{j_1}p^{i_2}_{j_2} g^{j_1 j_2}
\label{D2}
\end{equation}
\par
\df. A change of coordinates $(x)\to(\tilde x)$ (\ref{D1}) is called
isometric (conserves a metric tensor), if
$\t g_{ij}=g_{ij},\quad i,j=1,\ldots,n$, where $\t g_{ij}$ are defined by
the formula (\ref{D2}).\par

Every linear change of coordinates (\ref{D1}) corresponds to a linear
transformation of Euclidean space $L:\E\to\E$
\begin{equation}
L(e^i)=p^i_j e^j.
\label{L}
\end{equation}
And inversely, every linear transformation of Euclidean space
$L:\E\to\E$, defined by the formula (\ref{L}),
corresponds to the linear change of coordinates (\ref{D1}).
\par
\df.  A linear transformation (\ref{L}) of Euclidean space $\E$,
which corresponds to an isometric change of coordinates, is called
isometry.\par

If $L:\E\to\E$ is isometry, then
\begin{equation}
(L(U),L(V))=(U,V)\quad \forall U,V\in\E.
\label{LL}
\end{equation}
A set of all isometries is a group.
\par
\subsection{ Clifford algebra of Euclidean space.}
 We will use objects
$$
a^{i_1 i_2 \ldots i_k},\quad 1\leq i_1<i_2<\cdots<i_k\leq n,
$$
numbered by ordered multi-indices of the length  $k$ $(0\leq k\leq n)$.
 A number of all different ordered multi-indices of the length $k$
 is equal to a binomial coefficient
 $C^k_n=n!/k!(n-k)!$ .
And a number of all different multi-indices of the length from $0$ to
$n$ is equal to
$$
C^0_n+C^1_n+\cdots+C^n_n=2^n.
$$
 Now we come to the definition of Clifford algebra of Euclidean space.
Let $\E$ be an $n$-dimensional Euclidean space with a given metric tensor
$g$, with basis vectors  $e_1,\ldots,e_n$
 and corresponding basis covectors  $e^1,\ldots,e^n$ ( $e^i=g^{ij}e_j$).
 And let $\K$ be a field of real numbers $\R$, or complex numbers $\C$.
 One can consider a $2^n$-dimensional vector space $\S$ over the field
$\K$ with basis elements numbered by ordered multi-indices
\begin{equation}
e,e^i,e^{i_1 i_2},\ldots,e^{1\ldots n}\quad
1\leq i\leq n,\quad 1\leq i_1<i_2\leq n, \ldots.
\label{cl:basis}
\end{equation}
 Basis elements of the space $\S$, in particular, contain basis covectors
$e^1,\ldots,e^n$  of Euclidean space  $\E$ . So, $\E$ is a subspace of
the vector space $\S$. Let us define a multiplication of elements of the vector space $\S$ with the aid of the following rules:
\medskip

\noindent 1)$\quad(\alpha U)V=U(\alpha V)=\alpha(UV)$ ,
$\forall U,V\in {\cal S},\alpha\in\K$.

\noindent 2)$\quad(U+V)W=UW+VW,\quad W(U+V)=WU+WV$ ,  $\forall U,V,W\in {\cal S}$.

\noindent 3)$\quad(UV)W=U(VW)$ ,  $\forall U,V,W\in {\cal S}$.

\noindent 4)$\quad eU=Ue=U$ ,\  $\forall U\in {\cal S}$.

\noindent 5)$\quad e^i e^j+e^j e^i=2g^{ij}e$  ,  $i,j=1,\ldots,n$.

\noindent 6)$\quad e^{i_1}\ldots e^{i_k}=e^{i_1\ldots i_k}$ ,
  $1\leq i_1<\cdots<i_k\leq n$.

\medskip
 The rules 1)--4) are standard axioms of an associative algebra  with the scalar unit $e$. And only rules 5),6) reflect the peculiarity of the construction under consideration.

 Using these rules, one can compute a result of multiplication of arbitrary elements from the basis
 (\ref{cl:basis}) . Actually, in order to find an element of $\S$, which is equal to a multiplication
 $e^{i_1}\ldots e^{i_k}$
 with unordered indices, one can rearrange multipliers in it with the aid
of the rule 5). Doing this, one must taking into account, that multiplication
$e^i e^j$  for
$i>j$  gives two terms, one with  $ - e^j e^i$  and another with  $2g^{ij} e$ .
As a result, the multiplication  $e^{i_1}\ldots e^{i_k}$,
with the aid of rules 1)--5),
transforms into a sum of multiplications of elements $e^i$
with ordered indices
$$
e^{i_1}\ldots e^{i_k}=
\alpha e^{l_1}\ldots e^{l_p}+\beta e^{m_1}\ldots e^{m_q} + \ldots,
$$
 where $l_1<\cdots<l_p,\ m_1<\cdots<m_q,\ldots;\ \alpha,\beta,\ldots\in\K$ .
Now, in accordance with the rule 6), one may write
$$
e^{i_1}\ldots e^{i_k}=
\alpha e^{l_1\ldots l_p}+\beta e^{m_1\ldots m_q} + \ldots,
$$
 that means, in a right part there is a linear combination of basis elements
 (\ref{cl:basis}).
 Finally, to compute a result of a multiplication of two elements of the basis
 (\ref{cl:basis}) , one must write
$$
e^{i_1\ldots i_k} e^{j_1\ldots j_r}=e^{i_1}\ldots e^{i_k} e^{j_1}\ldots e^{j_r}
$$
 and use the previous reasoning.

\medskip
\noindent{\bf  Example  1}.  Let us compute a multiplication
\begin{eqnarray*}
e^{13}e^{234}&=&e^1 e^3 e^2 e^3 e^4 = e^1(-e^2 e^3+2g^{23}e)e^3 e^4\\
&=&-g^{33}e^1 e^2 e^4+2g^{23}e^1 e^3 e^4=-g^{33}e^{124}+2g^{23}e^{134}.
\end{eqnarray*}

 The vector space $\S$ over a field $\K$ with the defined multiplication of elements is called (a real or complex depended on $\K$) Clifford algebra
of Euclidean space $\E$ and denoted by  $\cl(\E)$.
 Basis elements $e^1,\ldots,e^n$  are called generators of Clifford algebra
and the basis  (\ref{cl:basis})  is called Clifford basis of
$\cl(\E)$.  If the matrix
$\|g^{ij}\|$ is diagonal with  $r$  pieces of  $+1$  and  $s$  pieces of
$-1$
 and $r+s=n$ , then a corresponding Clifford algebra is denoted by
 $\cl(r,s)$,  and by  $\cl(n)$  if  $s=0$.  Elements of Clifford algebra
 $\cl(\E)$  are called $\gamma$- numbers.

\medskip
\noindent {\bf  Note}.  Clifford algebra was invented in the year 1878 by the English mathematician W.~K.~Clifford \
\cite{clifford},  who called it
{\sl geometrical algebra}. We will use a term {\sl $\gamma$-number} short
of the term {\sl geometrical number}.
\medskip

\noindent{\bf  Example 2}.  A real Clifford algebra  $\cl(0,1)$
 with only one generator $e^1$ , for which ${(e^1)}^2=-e$ , is isomorphic to a field of complex numbers
 $\C$.
\medskip

\noindent{\bf  Example  3}.  A real Clifford algebra  $\cl(0,2)$  with generators
$e^1,e^2$ , for which ${(e^1)}^2={(e^2)}^2=-e$ , is isomorphic to
 an algebra of quaternions with imaginary units
$i=e^1,\ j=e^2,\ k=e^{12}$.
\medskip

 An important property of Clifford algebra $\cl(\E)$ is given by the formula
 $$
 (u_k e^k)^2= (g^{ij}u_i u_j)e,
 $$
that is the square of every element from $\E\subset\cl(\E)$ is a scalar.


\subsection{An exterior algebra of Euclidean space.}
Let's define an exterior multiplication of $\gamma$-numbers from
$\cl(\E)$, which will be denoted by the sign $\wedge$.
For generators $e^i$ of Clifford algebra let's assume
\footnote{The peculiarity of the construction under consideration
that we use Clifford multiplication instead of the usual tensor
multiplication.
}

\begin{equation}
e^{i_1}\wedge e^{i_2}\wedge\ldots\wedge e^{i_k}=e^{[i_1}e^{i_2}\ldots
e^{i_k]},
\label{A2}
\end{equation}
where square brackets denote an alternation of indices (with the
division over $k!$). In particular, we get from (\ref{A2}) the main relation
of Grassmann algebra $e^i\wedge e^j=-e^j\wedge e^i$.
\medskip
\par
\noindent{\bf Example 4.}
\begin{equation}
e^{i_1}\wedge e^{i_2}=\frac{1}{2}(e^{i_1}e^{i_2}-e^{i_2}e^{i_1})=
e^{i_1}e^{i_2}-g^{i_1 i_2}e,
\label{N2}
\end{equation}

\begin{eqnarray*}
e^{i_1}\wedge e^{i_2}\wedge e^{i_3}&=&
\frac{1}{6}(e^{i_1}e^{i_2}e^{i_3}+
e^{i_3}e^{i_1}e^{i_2}+e^{i_2}e^{i_3}e^{i_1}\\
&&-e^{i_2}e^{i_1}e^{i_3}-
e^{i_1}e^{i_3}e^{i_2}-e^{i_3}e^{i_2}e^{i_1})\\
&=&e^{i_1}e^{i_2}e^{i_3}-g^{i_2 i_3}e^{i_1}+
g^{i_1 i_3}e^{i_2}-g^{i_1 i_2}e^{i_3},
\end{eqnarray*}

\begin{eqnarray*}
e^{i_1}\wedge e^{i_2}\wedge e^{i_3}\wedge e^{i_4}&=&
\frac{1}{24}(e^{i_1}e^{i_2}e^{i_3}e^{i_4}+\ldots)\\
&=&e^{i_1}e^{i_2}e^{i_3}e^{i_4}-g^{i_3 i_4}e^{i_1}e^{i_2}
+g^{i_2 i_4}e^{i_1}e^{i_3}
-g^{i_2 i_3}e^{i_1}e^{i_4}\\
&&-g^{i_1 i_4}e^{i_2}e^{i_3}
+g^{i_1 i_3}e^{i_2}e^{i_4}-g^{i_1 i_2}e^{i_3}e^{i_4}\\
&&+(g^{i_1 i_4}g^{i_2 i_3}-g^{i_1 i_3}g^{i_2 i_4}+g^{i_1 i_2}g^{i_3 i_4})e
\end{eqnarray*}
\par
It can be checked, that the formula (\ref{A2}) is equivalent to the
following formula:
\begin{equation}
e^{i_1}\wedge\ldots\wedge e^{i_k}=e^{i_1}\ldots e^{i_k}+
\sum_{r=1}^{[\frac{k}{2}]} \frac{(-1)^r}{r!} Q^r(e^{i_1}\ldots e^{i_k}),
\label{A6}
\end{equation}
where
$$
Q(e^{i_1}\ldots e^{i_k})=
\sum_{1\leq p<q\leq k} (-1)^{q-p-1}
g^{i_p i_q}e^{i_1}\ldots\check{e^{i_p}}\ldots\check{e^{i_q}}\ldots e^{i_k}.
$$
The sign $\check{}$ above a factor $\check{e^{i_p}}$ means, that
the given factor in product is passed, $Q^r$ is $r$ times application
of the operation $Q$,
$[\frac{k}{2}]$ is an integer part of number $k/2$.
\par
Let's note, that the formula (\ref{A6}) can be accepted for definition
of an exterior multiplication of $\gamma$-numbers instead of the formula
(\ref{A2}).
\par
The formula (\ref{A6}) has a
remarkable property, it can be easily converted and allows to express Clifford products
of generators $e^i$ using exterior products of these generators. Namely
\begin{equation}
e^{i_1}\ldots e^{i_k}=e^{i_1}\wedge \ldots\wedge  e^{i_k}+
\sum_{r=1}^{[\frac{k}{2}]} \frac{1}{r!} Q^r(e^{i_1}\wedge\ldots\wedge e^{i_k}),
\label{A7}
\end{equation}
where
$$
Q(e^{i_1}\wedge\ldots\wedge e^{i_k})=
\sum_{1\leq p<q\leq k} (-1)^{q-p-1}
g^{i_p i_q}e^{i_1}\wedge\ldots\wedge
\check{e^{i_p}}\wedge\ldots\wedge\check{e^{i_q}}\wedge\ldots\wedge e^{i_k}.
$$
\medskip
\par
\noindent{\bf Example 5.} The formula (\ref{A7}), in particular, gives
\begin{equation}
e^{i_1}e^{i_2}=e^{i_1}\wedge e^{i_2}+g^{i_1 i_2}e,
\label{A8}
\end{equation}
$$
e^{i_1}e^{i_2}e^{i_3}=e^{i_1}\wedge e^{i_2}\wedge e^{i_3}
+g^{i_2 i_3}e^{i_1}-g^{i_1 i_3}e^{i_2}+g^{i_1 i_2}e^{i_3},
$$
\begin{eqnarray*}
e^{i_1}e^{i_2}e^{i_3}e^{i_4}&=&
e^{i_1}\wedge e^{i_2}\wedge e^{i_3}\wedge e^{i_4}
+g^{i_3 i_4}e^{i_1}\wedge e^{i_2}
-g^{i_2 i_4}e^{i_1}\wedge e^{i_3}\\
&&+g^{i_2 i_3}e^{i_1}\wedge e^{i_4}
+g^{i_1 i_4}e^{i_2}\wedge e^{i_3}
-g^{i_1 i_3}e^{i_2}\wedge e^{i_4}+g^{i_1 i_2}e^{i_3}\wedge e^{i_4}\\
&&+(g^{i_1 i_4}g^{i_2 i_3}-g^{i_1 i_3}g^{i_2 i_4}+g^{i_1 i_2}g^{i_3 i_4})e.
\end{eqnarray*}
\par
Combining the right and left parts of the formulas (\ref{A8})
$$
e^i e^j=e^i\wedge e^j+g^{ij}e,\quad e^j e^i=e^j\wedge e^i+g^{ji}e,
$$
and using equality $e^i\wedge e^j=-e^j\wedge e^i$, we obtain the main
relation of Clifford algebra
$$
e^i e^j+e^j e^i=2g^{ij}e.
$$
The formula (\ref{A7}) allows to express elements of Clifford basis
(\ref{cl:basis}) as a linear combinations of the following elements
which form a new basis of $\cl(\E)$, so called Grassmann basis
\begin{equation}
e,e^i,e^{i_1}\wedge e^{i_2},\ldots,e^{1}\wedge\ldots\wedge e^n,
\quad
1\leq i\leq n,\quad 1\leq i_1<i_2\leq n, \ldots.
\label{gr:basis}
\end{equation}
And, on the contrary, the formula (\ref{A6}) allows to express elements
of Grassmann basis (\ref{gr:basis}) as linear combinations of elements
of Clifford basis (\ref{cl:basis}).
\par
Now, one can find a result of exterior multiplication
\begin{equation}
e^{i_1\ldots i_p}\wedge e^{j_1\ldots j_q}
\label{A12}
\end{equation}
of any elements of Clifford basis. It is made in three steps.
\medskip
\par
G1. We express basis elements $e^{i_1\ldots i_p},e^{i_1\ldots i_q}$
using elements of Grassmann basis (\ref{gr:basis}) and substitute
the appropriate expressions in (\ref{A12}).
\par
G2. Further we compute how it is made for elements of Grassmann algebra,
representing outcome as a sum of basis elements (\ref{gr:basis}).
\par
G3. We write the result in Clifford basis with the aid of formulas
(\ref{A6}).
\medskip
\par
\noindent{\bf Example 6.}
\begin{eqnarray*}
e^{13}\wedge e^{23}&=&(e^1\wedge e^3+g^{13}e)\wedge(e^2\wedge e^3+g^{23}e)\\
&=&g^{23}e^1\wedge e^3+g^{13}e^2\wedge e^3+g^{13}g^{23}e\\
&=&g^{23}(e^{13}-g^{13}e)+g^{13}(e^{23}-g^{23}e)+g^{13}g^{23}e\\
&=&g^{23}e^{13}+g^{13}e^{23}-g^{13}g^{23}e.
\end{eqnarray*}
\par
Similarly one can find a result of Clifford multiplication of any elements
of Grassmann basis
\begin{equation}
(e^{i_1}\wedge\ldots\wedge e^{i_p})(e^{j_1}\wedge\ldots\wedge e^{j_q}),
\quad i_1<\cdots<i_p;\ j_1<\cdots<j_q.
\label{A13}
\end{equation}
It is made in three steps too.
\medskip
\par
C1. We express basis elements
$e^{i_1}\wedge\ldots\wedge e^{i_p}$, $e^{j_1}\wedge\ldots\wedge e^{j_q}$
using elements of Clifford basis (\ref{cl:basis}) and substitute the
appropriate expressions in (\ref{A13}).
\par
C2. Further we compute how it is made for elements of Clifford algebra,
representing outcome as a sum of basis elements (\ref{cl:basis}).
\par
C3. We write the result in Grassmann basis with the aid of formulas
(\ref{A7}).
\medskip
\par
\noindent{\bf Example 7.} It is easy to check up, that
$$
(e^1\wedge e^3)(e^2\wedge e^3)=-g^{33}e^1\wedge e^2+g^{23}e^1\wedge e^3-
g^{13}e^2\wedge e^3+(g^{13}g^{23}-g^{12}g^{33})e.
$$
\par
Let $t^\alpha, \alpha=1,\ldots,2^n$ be elements of Grassmann basis
(\ref{gr:basis}). According to the described algorithm, we can calculate
all Clifford products $t^\alpha t^\beta$ and write them down as linear
combinations of the elements of Grassmann basis (on $\gamma$ summation
from $1$ to $2^n$)
\begin{equation}
t^\alpha t^\beta=c^{\alpha\beta}_{\gamma} t^\gamma,\quad
\alpha,\beta=1,\ldots,2^n.
\label{tau}
\end{equation}
\par

Thus, we come to a $2^n$-dimensional vector space over the field $\K$
with two operations of multiplication -- Clifford multiplication
and an exterior multiplication and with two basises --
(\ref{cl:basis}) and (\ref{gr:basis}).
Both operations of multiplication satisfy axioms of associativity and
distributivity. The element $e$ is a scalar unit for both operations.
Let's denote this object by $\Lambda(\E)$ and call it an
exterior algebra of Euclidean space $\E$.
Elements of $\Lambda(\E)$ are called the forms. Basis covectors
$e^1,\ldots,e^n$ of Euclidean space $\E$, which contained
simultaneously in Clifford basis (\ref{cl:basis}) and in Grassmann basis
(\ref{gr:basis}) are called the generators  of an exterior algebra
$\Lambda(\E)$. If it is necessary to specify that a real or complex
algebra $\Lambda(\E)$ is considered, we shall write the appropriate index
$\R$ or $\C$, namely $\Lambda_{\R}(\E)$, $\Lambda_{\C}(\E)$.
\par
If the matrix $\|g^{ij}\|$ is diagonal, then each of the formulas
(\ref{A6}) and (\ref{A7}) gives the relations
\begin{equation}
e^{i_1}\ldots e^{i_k}=e^{i_1}\wedge\ldots\wedge e^{i_k}\quad
\hbox{when}\quad i_1<\cdots<i_k
\label{A15}
\end{equation}
meaning, that the basises (\ref{cl:basis}) and (\ref{gr:basis}) coincide.
In this simple case the indicated construction of $2^n$-dimensional
vector space with two operations of multiplication -- an exterior
multiplication and Clifford multiplication, was considered by many authors
(see for example, Rashevsky \cite{rashevsky}) first among them was
H.~Grassmann in the work of 1877 \cite{grassmann}.
He has considered an exterior algebra with two operation of multiplication--
an exterior multiplication and {\sl central} multiplication. The last one
has characteristic features of Clifford multiplication. With the help of
this construction Grassmann has tried to unit a calculus of exterior forms
and a calculus of quaternions. It is necessary to mean, that Clifford
algebra was invented in 1878, a year later Grassmann's paper. A detailed
controversy on this theme see in work of Doran, Hestenes etc. \cite{doran}.
The similar construction was considered for the case of arbitrary metric
in works of Pestov \cite{pestov}, Zhelnorovich \cite{zhelnorovich},
Benn and Tucher \cite{benn}, Lounesto \cite{lounesto} etc.
\par
\subsection{Main properties of the exterior algebra.}
Let $U$ be an arbitrary form from
$\LE$. It can be written as a decomposition over Grassmann basis
(\ref{gr:basis}) with coefficients from the field $\K$
\begin{equation}
U=ue+u_i e^i+\sum_{i_1<i_2}u_{i_1 i_2}e^{i_1}\wedge e^{i_2}+\ldots+
v_{1\ldots n}e^1\wedge\ldots\wedge e^n.
\label{B1}
\end{equation}
Let us assume, that coefficients  are antisymmetric with respect to all
indices
$$
u_{i_1\ldots i_k}=u_{[i_1\ldots i_k]},\quad i_1,\ldots,i_k=1,\ldots n.
$$
The forms
$$
\sum_{i_1<\ldots<i_k}u_{i_1\ldots i_k}e^{i_1}\wedge\ldots\wedge e^{i_k}=
\kfac u_{j_1\ldots j_k}e^{j_1}\wedge\ldots\wedge e^{j_k},\quad
u_{j_1\ldots j_k}=u_{[j_1\ldots j_k]}
$$
are called $k$-forms or rank $k$ forms. A rank $0$ forms
$ue$ are identified with the scalars $u$.  Let
$\Lambda^k(\E)$ be a vector subspace of all forms of rank $k$. Then
$$
\LE=\Lambda^0(\E)\oplus\Lambda^1(\E)\oplus\cdots\oplus\Lambda^n(\E)=
\Lambda^{\even}(\E)\oplus\Lambda^{\odd}(\E),
$$
where
$$
\Lambda^{\even}(\E)=\Lambda^0(\E)\oplus\Lambda^2(\E)\oplus\cdots,\quad
\Lambda^{\odd}(\E)=\Lambda^1(\E)\oplus\Lambda^3(\E)\oplus\cdots.
$$
The dimensions (real or complex depended on $\K$) of subspaces $\Lambda^k(\E)$
are equal to binomial coefficients $C^k_n$ and
$\sum^n_{k=0}C^k_n=2^n$. The dimensions of subspaces $\Lambda^{\even}(\E)$ and
$\Lambda^{\odd}(\E)$ are equal to  $2^{n-1}$. Elements of the space
$\Lambda^{\even}(\E)$ are called even forms and elements of $\Lambda^{\odd}(\E)$
are called odd forms. If $U\in\Lambda^r(\E)$,
$V\in\Lambda^s(\E)$, then
$U\wedge V\in\Lambda^{r+s}(\E)$ ($U\wedge V=0$, if $r+s>n$) and
$$
U\wedge V=(-1)^{rs}V\wedge U.
$$
This property is called anticommutativity.
In particular, if
$U,V\in\Lambda^{\even}(\E)$, then $U\wedge V\in\Lambda^{\even}(\E)$.
That means that $\Lambda^{\even}(\E)$ is a subalgebra of the exterior
algebra $\LE$.

Let us consider a linear change of coordinates
$(x)\to(\tilde{x})$ in Euclidean space $\E$, which is defined by
formulas (\ref{D1}),(\ref{D2}). Then, every form
$U\in\Lambda^k(\E)$ can be written in Grassmann basis (\ref{gr:basis})
and in new one
\begin{eqnarray*}
U&=&\kfac u_{i_1\ldots i_k} e^{i_1}\wedge\ldots\wedge e^{i_k}=
\kfac u_{i_1\ldots i_k}
(q^{i_1}_{j_1}\t e^{j_1})\wedge\ldots\wedge(q^{i_k}_{j_k}\t e^{j_k})\\
&=&\kfac q^{i_1}_{j_1}\ldots q^{i_k}_{j_k} u_{i_1\ldots i_k}
\t e^{j_1}\wedge\ldots\wedge\t e^{j_k}=
\kfac\t u_{j_1\ldots j_k}\t e^{j_1}\wedge\ldots\wedge\t e^{j_k},
\end{eqnarray*}
where
$$
\t u_{j_1\ldots j_k}=q^{i_1}_{j_1}\ldots q^{i_k}_{j_k} u_{i_1\ldots i_k},
\quad q^i_j=\frac{\partial x^i}{\partial\t x^j},\quad
u_{i_1\ldots i_k}=u_{[i_1\ldots i_k]},\quad
\tilde u_{j_1\ldots j_k}=\tilde u_{[j_1\ldots j_k]}
$$
Hence, according to the standard definitions of tensor analisys,
$U_{j_1\ldots j_k}$ is a covariant antisymmetric tensor of rank $k$.

\theorem 1. Let $U$ be an arbitrary form from  $\Lambda^k(\E)$
\begin{equation}
U=u_{i_1\ldots i_k} e^{i_1}\wedge\ldots\wedge e^{i_k},\quad
u_{i_1\ldots i_k}=u_{[i_1\ldots i_k]}.
\label{B2}
\end{equation}
Then
$$
U=\kfac u_{i_1\ldots i_k}e^{i_1}\ldots e^{i_k},
$$
where at right part there are Clifford multiplications of generators of
the algebra $\LE$.
\par
\proof. Is followed from the formula (\ref{A2})\fin

Let us introduce the following notations: $|g|$ is a module of
determinant of the matrix
$\|g_{ij}\|$; $\sgn(g)=\pm1$ is a sign of this determinant
$$
\det\|g_{ij}\|=1/\det\|g^{ij}\|=\sgn(g)|g|.
$$

\subsection{Hodge $\star$ operator.}
Vector spaces $\Lambda^k(\E)$ and $\Lambda^{n-k}(\E)$ have
equal dimensions $C^k_n=C^{n-k}_n$ and there is an operator which
identify forms from
$\Lambda^k(\E)$ with forms from $\Lambda^{n-k}(\E)$. If
$U\in\Lambda^k(\E)$ is written as (\ref{B2}), then
$\star U$ is the following form from $\Lambda^{n-k}(\E)$:
\begin{equation}
\star U=\frac{1}{k!(n-k)!} \sqrt{|g|}\,\epsilon_{i_1\ldots i_n}u^{i_1\ldots i_k}
e^{i_{k+1}}\wedge\ldots\wedge e^{i_n},
\label{star}
\end{equation}
where
$$
u^{i_1\ldots i_k}=g^{i_1 j_1}\ldots g^{i_k j_k}u_{j_1\ldots j_k}
$$
and $\epsilon_{i_1\ldots i_n}$ is a sign of the permutation of numbers
$(i_1\ldots i_n)$.

The form $\star U$ is a covariant antisymmetric tensor with respect to
a change of coordinates with a positive Jacobian. In this paper we shall
consider changes of coordinates only with positive Jacobian. Hence, we
do not distinguish tensors and pseudotensors.

The following formula is correct:
\begin{equation}
\star(\star U)=(-1)^{k(n+1)}\sgn(g)U,\quad U\in\Lambda^k(\E).
\label{A}
\end{equation}
In what follows, we establish a connection between $\star$ operator and
Clifford multiplication of  forms.
\medskip

We have considered the properties of forms which connected to the
exterior multiplication of forms. Now we begin to consider several
properties of forms connected to Clifford multiplication of forms.


\subsection{An operation of conjugation of forms.}
Let us suppose, that $U\in\Lambda^k(\E)$ is written as (\ref{B2}).
We may introduce operations $*$ and $rev$ (reverse) replacing
$e^{i_1}\wedge\ldots\wedge e^{i_k}$ by
$e^{i_k}\wedge\ldots\wedge e^{i_1}$
$$
U^{rev}=(-1)^{[\frac{k}{2}]}U=
\kfac u_{i_1\ldots i_k}e^{i_k}\wedge\ldots\wedge e^{i_1},
$$
$$
U^*=\kfac\bar u_{i_1\ldots i_k}e^{i_k}\wedge\ldots\wedge e^{i_1},
$$
where bar in the last formula means complex conjugation.
For $U,V\in\LE$
$$
(UV)^*=V^* U^*,\quad U^{**}=U,
$$
that means, the introduced operation $*$, which is called conjugation,
is an involution in $\LE$. Operations $*$ and $rev$ are identical for
$\K=\R$.


\subsection{A volume form.}
Let us introduce a form $I\in\Lambda^n(\E), \ (\dim\,\E=n)$
$$
I=\sqrt{|g|}\,e^1\wedge\ldots\wedge e^n=
\frac{1}{n!}\sqrt{|g|}\,\epsilon_{i_1\ldots i_n}
e^{i_1}\wedge\ldots\wedge e^{i_n}
$$
which is called the volume form (it plays an important role in a theory
of integration). It is easy to calculate, that
$$
II^*=I^*I=\sgn(g)e,\quad I^2=(-1)^{\frac{n(n-1)}{2}}\sgn(g)e.
$$
An important property of the volume form is expressed by the formula
$$
IU=(-1)^{k(n+1)}UI \quad\hbox{for}\quad \forall U\in\Lambda^k(\E).
$$
In other words, in case of even $n$ the form $I$ is commute with all
even forms from
$\Lambda^{\even}(\E)$ and anticommute with all odd forms from
$\Lambda^{\odd}(\E)$. In case of odd $n$ the form $I$ commute with all
forms from  $\Lambda(\E)$.

With the aid of volume form, the previously introduced Hodge $\star$
operator can be written as
\begin{equation}
\star U=U^{rev}I\quad\hbox{for}\quad\forall U\in\LE
\label{star1}
\end{equation}
Let's prove the correctness of the formula
(\ref{A}) with the aid of this relation. If
$U\in\Lambda^k(\E)$, then
$$
\star(\star U)=\star(U^{rev}I)=I^*UI=(-1)^{k(n+1)}\sgn(g)U.
$$


\subsection{A trace of form.}
\df. A trace of form is a linear operation
$\tr:\LE\to\K$ such, that
$\tr(e)=1$ and $\tr(e^{i_1}\wedge\ldots\wedge e^{i_k})=0$ for all $k\geq1$.
\par
\medskip

The main property of
$\tr$ operation is the following:
\begin{equation}
\tr(AB-BA)=0\quad\hbox{for}\quad\forall A,B\in\LE.
\label{tr}
\end{equation}
In particular, if $A=B^{-1}C$, then
$$
\tr(B^{-1}CB)=\tr\,C.
$$


\subsection{A structure of Euclidean space on $\LE$.}
Let's introduce a scalar multiplication of forms with the aid of the
following formula:
\begin{equation}
(U,V)=\tr(UV^*)\quad\hbox{for}\quad\forall U,V\in\LE.
\label{S}
\end{equation}
Such a scalar multiplication has properties
\smallskip

1. $ (U,V)=\overline{(V,U)}\quad\hbox{for}\quad\forall U,V\in\LE$.

2. $ \alpha(U,V)=(\alpha U,V)=(U,\bar{\alpha}V)
\quad\hbox{for}\quad\forall U,V\in\LE,\ \forall\alpha\in\K.$
\smallskip

The scalar multiplication (\ref{S}) gives us a structure of Euclidean
space on $\LE$ with a metric tensor $\hat g$ (it is called an exterior
metric tensor), which components are defined by the formulas
$$
\hat g^{\alpha\beta}=\hat g^{\beta\alpha}=(t^\alpha,t^\beta),\quad
\alpha,\beta=1,\ldots,2^n,
$$
where $t^\alpha,\ \alpha=1,\ldots,2^n$ are elements of Grassmann basis
(\ref{gr:basis}). The following formulas give us information about
the exterior metric tensor $\hat g$:
\begin{equation}
(e^{i_1}\ww e^{i_k})(e^{i_1}\ww e^{i_k})^*=
(-1)^{\frac{k(k-1)}{2}}(e^{i_1}\ww e^{i_k})^2=
M^{i_1\ldots i_k}_{i_1\ldots i_k}e,
\label{5*}
\end{equation}
\begin{equation}
(e^{i_1}\ww e^{i_k},e^{j_1}\ww e^{j_k})=M^{i_1\ldots i_k}_{j_1\ldots j_k},
\label{5**}
\end{equation}
\begin{equation}
(e^{i_1}\ww e^{i_k},e^{j_1}\ww e^{j_r})=0\quad\hbox{¯à¨}\quad k\neq r.
\label{5***}
\end{equation}
Here $M^{i_1\ldots i_k}_{j_1\ldots j_k}$ is a minor of the matrix
$\|g^{ij}\|$, that is a determinant of the matrix composed from elements
of the matrix
$\|g^{ij}\|$ standing on the intersection of lines with numbers
$i_1\ldots
i_k$ and columns with numbers $j_1\ldots j_k$.
\par

\medbreak
\noindent{\bf Examples.}
$$
(e^{i_1}\wedge e^{i_2})^2=-(g^{i_1 i_1}g^{i_2 i_2}-(g^{i_1 i_2})^2)e,
$$
$$
(e^1\ww e^n)^2=(-1)^{\frac{n(n-1)}{2}}\det\|g^{ij}\|e,
$$
$$
(e^{i_1}\wedge e^{i_2},e^{j_1}\wedge e^{j_2})=
\det\left\|
\begin{array}{cc}
g^{i_1 j_1} & g^{i_1 j_2}\\
g^{i_2 j_1} & g^{i_2 j_2}
\end{array}
\right\|=g^{i_1 j_1}g^{i_2 j_2}-g^{i_2 j_1}g^{i_1 j_2}.
$$
From the formulas
(\ref{5*}),(\ref{5**}),(\ref{5***}) it is evident, that the matrix
$\|\hat g^{\alpha\beta}\|$ of exterior metric tensor has the dimension
$2^n\times 2^n$ and is block diagonal with square blocks
$\hat g_{(k)}$ of the dimensions
$C^k_n\times C^k_n$ and
$$
\hat g_{(0)}=1,\quad \hat g_{(1)}=\|g^{ij}\|,\quad\hat
g_{(n)}=\det\|g^{ij}\|.
$$
The matrix $\hat g_{(k)}$ is called $k$-associated matrix to the matrix
$\|g^{ij}\|$ (\cite{gantmacher}). Hence, not only vector space
$\LE$ becomes Euclidean space with the metric tensor
$\hat g$, but also all of subspaces
$\Lambda^k(\E),\ k=0,\ldots,n$ become Euclidean spaces with the
corresponding metric tensors $\hat g_{(k)}$.

In particular, a vector space
$\Lambda^1(\E)$ with the metric tensor $\hat g_{(1)}$ becomes
Euclidean space
and also $\Lambda^1_\R(\E)$ with the metric tensor $\hat g_{(1)}$ becomes
Euclidean space
which is isomorphic to initial Euclidean space
$\E$ with the metric tensor $g$.

\remark\, 1. If the $n\times n$-matrix $\|g^{ij}\|$ of the metric tensor
of $\E$ is diagonal, then the
$2^n\!\times\!2^n$-matrix
$\|\hat g^{\alpha\beta}\|$-matrix of the exterior metric tensor of
$\LE$, is also diagonal.

\remark\, 2. If the matrix $\|g^{ij}\|$ is positive defined, then
the matrix $\|\hat g^{\alpha\beta}\|$ is also positive defined.


\subsection{Lie algebra of  $2$-forms.}
Let us introduce an operation of commutation of $2$-forms
$$
\com(U,V)=[U,V]=UV-VU,\quad U,V\in\Lambda^2(\E),
$$
in which Clifford multiplication of forms is used. It is easy to check,
that for basis $2$-forms there is the formula
\begin{equation}
\com(e^{i_1}\wedge e^{i_2},e^{j_1}\wedge e^{j_2})=
-2g^{i_1 j_1}e^{i_2}\wedge e^{j_2}-2g^{i_2 j_2}e^{i_1}\wedge e^{j_1}
+2g^{i_1 j_2}e^{i_2}\wedge e^{j_1}+2g^{i_2 j_1}e^{i_1}\wedge e^{j_2}
\label{com}
\end{equation}
Therefore, a set of $2$-forms
$\Lambda^2(\E)$ is closed with respect to the operation of commutation.
For this operation a Jacobi identity is satisfied and, consequently, the
set of $2$-forms with operation of commutation is a Lie algebra.

\theorem 2.
\begin{enumerate}
\item For the space dimensions
$n\leq 3$ a Clifford multiplication of forms can be expressed
with the aid of exterior multiplication and Hodge $\star$ operation.
\item For $n=4,5$  a Clifford multiplication of forms can be expressed
with the aid of exterior multiplication, Hodge $\star$ operation and
bilinear operation
$\com\,:\,\Lambda^2(\E)\times\Lambda^2(\E)\to\Lambda^2(\E)$,
which is defined by the formula (\ref{com}).
\end{enumerate}
\par

\proof. Let us write an explicit formulas for $n=2,3,4$ in which ranks
of forms indicated in the designation
$\s{k}{U}\in\LkE$. First of all let us note, that for all
$n\geq1$
$$
\s{0}{U}\s{k}{V}=\s{k}{V}\s{0}{U}=\s{0}{U}\wedge\s{k}{V}
=\s{k}{V}\wedge\s{0}{U},\quad k=0,1,\ldots,n.
$$
\noindent For $n=2$
\begin{eqnarray*}
\s{1}{U}\s{1}{V}&=&\s{1}{U}\wedge\s{1}{V}+
\star(\s{1}{U}\wedge\star\s{1}{V})\sgn(g),\\
\s{1}{U}\s{2}{V}&=&\star \s{1}{U} \wedge  \star \s{2}{V} \sgn(g),\\
\s{2}{U}\s{1}{V}&=&-\star \s{2}{U} \wedge  \star \s{1}{V} \sgn(g),\\
\s{2}{U}\s{2}{V}&=&-\star \s{2}{U} \wedge  \star \s{2}{V} \sgn(g).
\end{eqnarray*}

\noindent Forì $n=3$
\begin{eqnarray*}
\s{1}{U}\s{1}{V}&=&\s{1}{U}\wedge\s{1}{V}+\star(\s{1}{U}\wedge\star\s{1}{V})\sgn(g),\\
\s{1}{U}\s{2}{V}&=&\s{1}{U}\wedge\s{2}{V}-\star(\s{1}{U}\wedge\star\s{2}{V})\sgn(g),\\
\s{1}{U}\s{3}{V}&=&\star \s{1}{U} \wedge  \star \s{3}{V} \sgn(g),\\
\s{2}{U}\s{1}{V}&=&\s{2}{U}\wedge\s{1}{V}-\star(\star\s{2}{U}\wedge\s{1}{V})\sgn(g),\\
\s{2}{U}\s{2}{V}&=&-\star\s{2}{U}\wedge\star\s{2}{V}\sgn(g)-\star(\s{2}{U} \wedge  \star \s{2}{V})\sgn(g),\\
\s{2}{U}\s{3}{V}&=&-\star \s{2}{U} \wedge  \s{3}{V} \sgn(g),\\
\s{3}{U}\s{1}{V}&=&\star \s{3}{U} \wedge  \star \s{1}{V} \sgn(g),\\
\s{3}{U}\s{2}{V}&=&-\star \s{3}{U} \wedge  \star \s{2}{V} \sgn(g),\\
\s{3}{U}\s{3}{V}&=&-\star \s{3}{U} \wedge  \star \s{3}{V} \sgn(g).
\end{eqnarray*}

\noindent For $n=4$
\begin{eqnarray*}
\s{1}{U}\s{1}{V}&=&\s{1}{U} \wedge  \s{1}{V}+\star (\s{1}{U} \wedge  \star \s{1}{V})\sgn(g),\\
\s{1}{U}\s{2}{V}&=&\s{1}{U} \wedge  \s{2}{V}+\star (\s{1}{U} \wedge  \star \s{2}{V})\sgn(g),\\
\s{1}{U}\s{3}{V}&=&\s{1}{U} \wedge  \s{3}{V}+\star (\s{1}{U} \wedge  \star \s{3}{V})\sgn(g),\\
\s{1}{U}\s{4}{V}&=&\star \s{1}{U} \wedge  \star \s{4}{V} \sgn(g),\\
\s{2}{U}\s{1}{V}&=&\s{2}{U} \wedge  \s{1}{V}-\star (\star \s{2}{U} \wedge  \s{1}{V})\sgn(g),\\
\s{2}{U}\s{2}{V}&=&\s{2}{U}\wedge\s{2}{V}-
\star(\s{2}{U}\wedge\star\s{2}{V})\sgn(g)+\frac{1}{2}\com(\s{2}{U},\s{2}{V}),\\
\s{2}{U}\s{3}{V}&=&-\star \s{2}{U} \wedge  \star \s{3}{V} \sgn(g)+\star (\s{2}{U} \wedge  \star \s{3}{V})\sgn(g),\\
\s{2}{U}\s{4}{V}&=&-\star \s{2}{U} \wedge  \star \s{4}{V} \sgn(g),\\
\s{3}{U}\s{1}{V}&=&\s{3}{U} \wedge  \s{1}{V}-\star (\star \s{3}{U} \wedge  \s{1}{V})\sgn(g),\\
\s{3}{U}\s{2}{V}&=&\star \s{3}{U} \wedge  \star \s{2}{V} \sgn(g)+\star (\star \s{3}{U} \wedge  \s{2}{V})\sgn(g),\\
\s{3}{U}\s{3}{V}&=&-\star \s{3}{U} \wedge  \star \s{3}{V} \sgn(g)-\star (\s{3}{U} \wedge  \star \s{3}{V})\sgn(g),\\
\s{3}{U}\s{4}{V}&=&-\star \s{3}{U} \wedge  \star \s{4}{V} \sgn(g),\\
\s{4}{U}\s{1}{V}&=&-\star \s{4}{U} \wedge  \star \s{1}{V} \sgn(g),\\
\s{4}{U}\s{2}{V}&=&-\star \s{4}{U} \wedge  \star \s{2}{V} \sgn(g),\\
\s{4}{U}\s{3}{V}&=&\star \s{4}{U} \wedge  \star \s{3}{V} \sgn(g),\\
\s{4}{U}\s{4}{V}&=&\star \s{4}{U} \wedge  \star \s{4}{V} \sgn(g)\fin
\end{eqnarray*}

In particular, from the above formulas we get the formula, which will be
used in what follows
\begin{eqnarray}
\s{1}{U}\s{k}{V}&=&\s{1}{U}\wedge\s{k}{V}+
(-1)^{k+1}\star^{-1}(\s{1}{U}\wedge\star\s{k}{V})\label{U1Vk}\\
&=&\s{1}{U}\wedge\s{k}{V}+
(-1)^{n(k+1)}\star(\s{1}{U}\wedge\star\s{k}{V})\sgn(g),\
k=0,1,\ldots,n\nonumber
\end{eqnarray}

Theorem 2 for $\dim(\E)\leq4$ gives a new method of introduction of
Clifford multiplication for elements of Grassmann algebra.


\subsection{A group $\Spin(\E)$.}
Let $F$ be an arbitrary fixed form from $\LE$.
One can consider an operator $L_F\,:\,\LE\to\LE$ such that
\begin{equation}
L_F(U)=F^*UF\quad\hbox{for}\quad U\in\LE.
\label{L_F}
\end{equation}
It is evident, that $L_F(U)=L_{-F}(U)$, that means two forms $+F,-F$
from $\LE$ generate the same operator $L_F$.

Let us introduce the following set of forms, playing an exclusive role in
our construction:
\begin{equation}
\Spin(\E)=
\{F\in\Lambda^\even_\R(\E)\,:\,FF^*=e,\
L_F\,:\,\Lambda^1(\E)\to\Lambda^1(\E)\}.
\label{Spin}
\end{equation}
This set is a group with respect to Clifford multiplication. It is
called a spinor group of Euclidean space $\E$.

\theorem 3 (\cite{benn}). If $n\leq5$, and
$F\in\Lambda^\even_\R(\E),\ FF^*=e$,
then
$$
L_F\,:\,\Lambda^1(\E)\to\Lambda^1(\E).
$$
\par
In \cite{benn} there is an example which proves that the theorem is not
true for
$n\geq6$\fin
\medskip

It follows from the theorem 3, that for $n\leq5$ the definition of
group $\Spin(\E)$ is simpler
\begin{equation}
\Spin(\E)=
\{F\in\Lambda^\even_\R(\E)\,:\,FF^*=e\}\quad\hbox{for}\quad n\leq5.
\label{Spin5}
\end{equation}

\theorem 4. If $F\in\Spin(\E)$, then
$$
L_F\,:\,\LkE\to\LkE,\quad k=0,1,\ldots,n.
$$
\par

\proof. Let $F\in\Spin(\E)$, $F^* e^i F=p^i_j e^j$ and
$U\in\Lambda^k(\E)$. From the theorem 1 we get
$$
U=\kfac u_{i_1\ldots i_k}e^{i_1}\ww e^{i_k}=\kfac u_{i_1\ldots i_k}e^{i_1}\ldots
e^{i_k},\quad u_{i_1\ldots i_k}=u_{[i_1\ldots i_k]}.
$$
Therefore
\begin{eqnarray*}
L_F(U)&=&F^*UF=\kfac u_{i_1\ldots i_k}(F^*e^{i_1}F)\ldots(F^*e^{i_k}F)=
\kfac p^{i_1}_{j_1}\ldots p^{i_k}_{j_k}u_{i_1\ldots i_k}e^{j_1}\ldots e^{j_k}\\
&=&\kfac u^\prime_{j_1\ldots j_k}e^{j_1}\ldots e^{j_k}=
\kfac u^\prime_{j_1\ldots j_k}e^{j_1}\ww e^{j_k}.
\end{eqnarray*}
According to the theorem 1, the last equality is true since
$u^\prime_{j_1\ldots j_k}=u^\prime_{[j_1\ldots j_k]}$\fin


\subsection{Isometries and Spin-isometries of the space $\LE$.}

\theorem 5. If $F\in\Spin(\E)$, then the operator
$L_F\,:\,\LkE\to\LkE,\ k=0,1,\ldots,n$ is an isometry of Euclidean space
$\LE$ with the scalar multiplication
(\ref{S}), and also it is isometry in every Euclidean space $\LkE$ with
(\ref{S}).
\par

\proof. Let $U,V\in\LkE$, or $U,V\in\LE$. Then
\begin{eqnarray*}
(L_F(U),L_F(V))&=&\tr(L_F(U)(L_F(V))^*)=\tr(F^*UFF^*V^*F)\\
&=&\tr(F^*(UV^*)F)=\tr(UV^*)=(U,V)\fin
\end{eqnarray*}
\medskip

\noindent{\bf Consequence.} Since Euclidean space $\Lambda^1_\R(\E)$
with the scalar multiplication (\ref{S}) is isomorphic to initial
Euclidean space $\E$, then for all
$F\in\Spin(\E)$ the operator
$L_F\,:\,\Lambda^1_\R(\E)\to\Lambda^1_\R(\E)$ written in the form
$$
L_F(e^i)=a^i_j(F)e^j,
$$
where $a^i_j(F)$ are real numbers, define an isometry of Euclidean space
$\E$.

\df. A linear transformation $L\,:\,\E\to\E$ of Euclidean space $\E$
\begin{equation}
L(e^i)=p^i_j e^j,\quad i=1,\ldots,n
\label{SI}
\end{equation}
defined by real numbers $p^i_j,\ i,j=1,\ldots,n$, is called
Spin-isometry of Euclidean space $\E$, if there exists a form
$F\in\Spin(\E)$ (a pair of forms $+F,-F$) such, that
$$
p^i_j e^j=F^*e^i F,\quad i=1,\ldots,n
$$
and this equality must be considered as an equality of two forms
from  $\LE$
(in other words $L(e^i)=L_F(e^i)$).
\par
\medskip

The set of all Spin-isometries of $\E$ is a subgroup of the group of all
isometries of $\E$.

Let us choose in Euclidean space $\E$ an orthonormal basis
$e^1,\ldots,e^n$ in which the matrix of metric tensor $g$ is diagonal
with  $r$ pieces of $+1$ and $s$ pieces of $-1$ and
$r+s=n$. In this case the corresponding spinor group is denoted by
$\Spin(r,s)$,  and $\Spin(n)$ if $s=0$. Since for every
$F\in\Spin(r,s)$ the map $L_F\,:\,\E\to\E$ is isometry (conserves the
metric), then we get a homomorphism
$$
f\ :\ \Spin(r,s)\to O(r,s).
$$
The kernel of homomorphism consists of $+1$ and $-1$. The range of it
depends on numbers $r$ and $s$. If
$s=0$, then $O(r,s)=O(n)$ and $f(\Spin(n))=SO(n)$. That means
$$
\Spin(n)/\{\pm1\}=SO(n).
$$
For $r>0, s>0$ we have $f(\Spin(r,s))=SO^+(r,s)$, that means
$$
\Spin(r,s)/\{\pm1\}=SO^+(r,s),
$$
where $SO^+(r,s)$ is a connected component of unity element of the group
$SO(r,s)$.

In particular case $r=1,\ s=3$ the group $\Spin(1,3)$ is isomorphic to
the group
$SL(2,\C)$.

\df. The linear change of coordinates of Euclidean space $\E$ (\ref{D1})
\begin{equation}
\t x^i=p^i_j x^j,\quad \t e^i=p^i_j e^j
\label{zam1}
\end{equation}
is called Spin-isometric (Si-change of coordinates), if there exists
a form $F\in\Spin(\E)$ such, that
\begin{equation}
\t e^i=p^i_j e^j=F^*e^i F.
\label{zam2}
\end{equation}
In this case we say, that Si-change of coordinates (\ref{zam2})
is associated with the form $F\in\Spin(\E)$.
\medskip

Si-change of coordinates (\ref{zam2}) of Euclidean space
$\E$ corresponds to Spin-isometry by the formula (\ref{SI}) and vice
verse.

Let us consider Si-change of coordinates (\ref{zam1}),(\ref{zam2})
associated with a form
 $F\in\Spin(\E)$.

\theorem 6. (The first property of a  Si-change of coordinates.)
In a new system of coordinates $(\tilde x)$ the form (\ref{B2})
can be written as
$$
U=F^*(\sum^n_{k=0}\kfac u_{i_1\ldots i_k}\t e^{i_1}\ww \t e^{i_k})F.
$$
\par

\proof. The theorem 1 gives
\begin{eqnarray*}
U&=&\kfac u_{i_1\ldots i_k} e^{i_1}\ww e^{i_k}=
\kfac u_{i_1\ldots i_k} e^{i_1}\ldots e^{i_k}\\
&=&\kfac u_{i_1\ldots i_k}(F\t e^{i_1}F^*)\ldots(F\t e^{i_k}F^*)=
F(\kfac u_{i_1\ldots i_k}\t e^{i_1}\ldots \t e^{i_k})F^*\\
&=&F(\kfac u_{i_1\ldots i_k}\t e^{i_1}\ww \t e^{i_k})F^*\fin
\end{eqnarray*}

If there is a change of coordinates $(x)\to(\t x)$, then the value
$J=\det\|\frac{\partial x^i}{\partial \t x^j}\|$
is called Jacobian of this change of coordinates.

\theorem 7. (The second property of  Si-change of coordinates.)
For $n\leq4$ Jacobian of  Si-change of coordinates is equal to one.
\par

\proof. Since for a Si-change of coordinates the metric tensor is
conserved, then
$$
\det\|\t g_{ij}\|=
\det\|g_{kl}\frac{\partial x^k}{\partial\t x^i}\frac{\partial x^l}{\partial\t
x^j}\|=J^2 \det\|g_{ij}\|.
$$
That means $J=\pm1$. For $n\leq4$ we have checked, that $J=1$\fin
\medskip

A consideration of the exterior algebra $\LE$ is finished.

\vfill\eject
\section{Exterior differential forms on Riemannian manifold.}

\subsection{An elementary Riemannian manifold $\V$.}
Let $M$ be an $n$-dimensional differentiable manifold covered by a system
of coordinates
$x^1,\ldots,x^n$. We  shall consider atlases on  $M$ atlases each of which
consists of one chart. Let us suppose, that on $M$ there is a smooth
twice covariant tensor field (a metric tensor) which in coordinates
$x^1,\ldots,x^n$ has components
$g_{ij}=g_{ij}(x),\ x\in M$ such that

1. $g_{ij}=g_{ji},\ i,j=1,\ldots n$.

2. The matrix $\|g_{ij}\|$ is nondegenerate for all $x\in M$.
\smallskip

The full set of $\{M,g_{ij}\}$ is called manifold  $\V$. The manifold
$\V$ is an elementary Riemannian \footnote{We do not require the
positive definiteness of the matrix $\|g_{ij}\|$.} manifold (with one chat
atlases).

The tensor field
$g_{ij}$ convert each of tangent spaces
$T_x$ to the manifold $\V$ in a point $x$ into Euclidean space
$\E_x$  with scalar multiplication
$$
(u,v)=g_{ij}(x)u^i v^j,\quad u,v\in T_x.
$$
Let us denote by $\LkV,\ k=0,1,\ldots n$ the sets of exterior differential
forms on
$\V$ (covariant antisymmetric tensor fields) of rank $k$
over a field $\K$  of real or complex numbers. Elements of $\LkV$  are
called
$k$-forms. Each $k$-form can be written as
\be
U=\kfac u_{j_1\ldots j_k}dx^{j_1}\ww dx^{j_k}=
\sum_{i_1<\cdots<i_k} u_{i_1\ldots i_k}dx^{i_1}\ww dx^{i_k},
\label{kform}
\ee
where $u_{j_1\ldots j_k}=u_{[j_1\ldots j_k]}$ --  antisymmetric, with
respect to all indices, functions of
$x\in\V$ taking its values in the field $\K$. Then
$$
\LV=\Lambda^0(\V)\oplus\Lambda^1(\V)\oplus\cdots\oplus\Lambda^n(\V)=
\Lambda^{\even}(\V)\oplus\Lambda^{\odd}(\V).
$$
Elements of $\LV$ (forms), can be written as linear combinations of
$2^n$ basis forms
\begin{equation}
1,dx^i,dx^{i_1}\wedge dx^{i_2},\ldots,dx^{1}\wedge\ldots\wedge dx^n,
\quad\hbox{£¤¥}\quad
1\leq i\leq n,\quad 1\leq i_1<i_2\leq n, \ldots.
\label{dx:basis}
\end{equation}
with coefficients from $\K$, which depend on $x\in\V$.
For elements $\LV$ an exterior multiplication is defined, and  if
$U\in\Lambda^r(\V),V\in\Lambda^s(\V)$, then
$$
U\wedge V=(-1)^{rs}V\wedge U\in\Lambda^{r+s}(\V)
$$
and $U\wedge V=0$ for $r+s>n$.
The form
$$
I=\sqrt{|g|}\,dx^1\ww dx^n
$$
is called a volume form.
Let $t^\alpha,\ \alpha=1,\ldots,2^n$ are basis forms from (\ref{dx:basis}).
We may define Clifford multiplication of forms from $\LV$ as a bilinear
operation
$\LV\times\LV\to\LV$, which for basis forms (\ref{dx:basis}) calculated
with the aid of rules C1,C2,C3 of chapter 1.4 (where we must replace $e^i$
by $dx^i$).
\medskip

\example. Let us consider an elementary Riemannian manifold $\V$
of the dimension $n=2$ with coordinates
$x^1,x^2$ and with a metric tensor, defined by its contravariant
components
$g^{11},g^{12}=g^{21},g^{22}$. Basis forms $t^1,t^2,t^3,t^4$ are
$$
1,\ dx^1,\ dx^2,\ dx^1\wedge dx^2.
$$
Then, Clifford multiplication is defined with the aid of formulas
\begin{eqnarray*}
(dx^1)^2&=&g^{11},\\
(dx^2)^2&=&g^{22},\\
dx^1 dx^2&=&dx^1\wedge dx^2+g^{12},\\
dx^2 dx^1&=&-dx^1\wedge dx^2+g^{12},\\
dx^1(dx^1\wedge dx^2)&=&-(dx^1\wedge dx^2)dx^1=-g^{12}dx^1+g^{11}dx^2,\\
dx^2(dx^1\wedge dx^2)&=&-(dx^1\wedge dx^2)dx^2=-g^{22}dx^1+g^{12}dx^2,\\
(dx^1\wedge dx^2)(dx^1\wedge dx^2)&=&(g^{12})^2-g^{11}g^{22}
\end{eqnarray*}
and $1t^j=t^j1=t^j,\ j=1,2,3,4$. In particular, from these relations
imply the formula
$dx^1 dx^2+dx^2 dx^1=2 g^{12}$.
\bigskip

The definitions of operators $*$, $\star$,
$\tr$, $\exp$ automatically transfer from $\LE$ to
$\LV$.


\subsection{Groups $\UV$ and $\Spin(\V)$.}
Let us introduce a set of smooth complex functions on $\V$ ($0$-forms)
with absolute value equal to one for every $x\in\V$
$$
\UV=\{a(x)\in\Lambda^0_\C(\V)\,:\,|a(x)|=1, \ x\in\V\}.
$$
This set is a group with respect to multiplication. The set of functions
from $\UV$, considering in the fixed point $x\in\V$, form a Lie group
${\rm U}(1)$.

Also, let us consider the following set of forms from $\LV$:
$$
\Spin(\V)=
\{F\in\Lambda^\even_\R(\V)\,:\,FF^*=1,\
L_F\,:\,\Lambda^1(\V)\to\Lambda^1(\V)\},
$$
where an operator $L_F$ is defined by the formula (\ref{L_F}).
This set of forms is a group with respect to Clifford multiplication.
Forms from $\Spin(\V)$, considering in the fixed point
$x\in\V$, compose a group $\Spin(\E_x)$,
where $\E_x$ is the tangent Euclidean space to Riemannian manifold
$\V$ in the point $x$.


\subsection{Spin-isometric change of coordinates on $\V$.}
Let us consider a change of coordinates $(x)\to(\t x)$ on the manifold $\V$,
with  smooth functions
\begin{equation}
x^i=x^i(\t x^1,\ldots,\t x^n),\quad \t x^i=\t x^i(x^1,\ldots,x^n).
\label{xtx}
\end{equation}

\df. A change of coordinates (\ref{xtx}) on the manifold $\V$ is called
Spin-isometric (Si-change of coordinates), if there exists a form
$F\in\Spin(\V)$ (a pair of forms $+F,-F$) such, that
\begin{equation}
d\t x^i=\frac{\partial\t x^i}{\partial x^j}dx^j=F^*dx^i F.
\label{Si:rule}
\end{equation}
In this case we  shall say, that Si-change of coordinates (\ref{xtx})
is associated with the form $F\in\Spin(\V)$.
\medskip

\theorem 8. (The first property of Si-change of coordinates).
Let a form $U\in\LkV$ is written in coordinates $x^1,\ldots,x^n$ as
$$
U=\kfac u_{i_1\ldots i_k}dx^{i_1}\ww dx^{i_k},\quad
u_{i_1\ldots i_k}=u_{[i_1\ldots i_k]}.
$$
and there is Si-change of coordinates $(x)\to(\t x)$ associated with
some form
$F\in\Spin(\V)$. Then, in coordinates $\t x^1,\ldots,\t x^n$
$$
U=\kfac u_{i_1\ldots i_k}\frac{\partial x^{i_1}}{\partial\t x^{j_1}}\ldots
\frac{\partial x^{i_k}}{\partial\t x^{j_k}}d\t x^{j_1}\ww d\t x^{j_k}=
F(\kfac u_{i_1\ldots i_k}d\t x^{j_1}\ww d\t x^{j_k})F^*.
$$
\par

\proof\,\, word for word repeats the proof of the theorem 6 (we must
replace
$e^i,\t e^i$ by $dx^i,d\t x^i$)\fin

\theorem 9. (The second property of Si-change of coordinates).
For $n\leq4$ Jacobian of  Si-change of coordinates is equal to one
for all $x\in\V$.
\par

\proof\,\, repeats the proof of the theorem 7\fin
\medskip


\subsection{Tensors with values in $\LkV$.}
Let in coordinates $(x^1,\ldots,x^n)$
$$
u^{m_1\ldots m_r}_{i_1\ldots i_k j_1\ldots j_s}(x)
=u^{m_1\ldots m_r}_{[i_1\ldots i_k]j_1\ldots j_s}(x),\quad x\in\V
$$
are components of a tensor field of the rank
$(r,k+s)$ antisymmetric with respect to the first
$k$ covariant indices. One may consider the following objects:
\be
U^{m_1\ldots m_r}_{j_1\ldots j_s}=
\kfac u^{m_1\ldots m_r}_{i_1\ldots i_k j_1\ldots
j_s}(x)\,dx^{i_1}\wedge\ldots\wedge dx^{i_k}
\label{index:form}
\ee
which are formally written as
$k$-forms. Under a change of coordinates
$(x)\to(\tilde x)$ the values
(\ref{index:form}) transform as components of tensor field of
rank $(r,s)$
\be
\tilde U^{a_1\ldots a_r}_{b_1\ldots b_s}=
q^{j_1}_{b_1}\ldots q^{j_s}_{b_s} p^{a_1}_{m_1}\ldots p^{a_r}_{m_r}
U^{m_1\ldots m_r}_{j_1\ldots
j_s},\quad q^j_b=\frac{\partial x^j}{\partial\tilde x^b},
\quad p^a_m=\frac{\partial\tilde x^a}{\partial x^m}.
\label{index:form1}
\ee
The objects (\ref{index:form}) are called tensors of rank $(r,s)$ with
values in $\LkV$. We shall write them as
$$
U^{m_1\ldots m_r}_{j_1\ldots j_s}\in\top^r_s\LkV.
$$
For $U_j\in\top_1\LkV$ we have
$$
dx^j U_j\in\Lambda^{k+1}(\V)\oplus\Lambda^{k-1}(\V).
$$


\subsection{Operators of covariant differentiation $\nabla_i$.}
It is known, that on Riemannian manifold Christoffel symbols
$\Gamma^k_{ij}=\Gamma^k_{ji}$ (its also
called Levi-Chivita connectedness components)
can be defined with the aid of metric
tensor by the formula
\begin{equation}
\Gamma^k_{ij}=\frac{1}{2}g^{kl}(\frac{\partial g_{lj}}{\partial x^i}
+\frac{\partial g_{il}}{\partial x^j}-\frac{\partial g_{ij}}{\partial
x^k}).
\label{chris}
\end{equation}
If on the manifold $\V$ a smooth change of coordinates (\ref{xtx}) is
defined, then Christoffel symbols
$\t\Gamma^c_{ab}$ in coordinates $\t x^1,\ldots,\t x^n$ relate with
Christoffel symbols in coordinates $x^1,\ldots,x^n$
$\Gamma^k_{ij}$ by the following relations:
\begin{equation}
\t\Gamma^c_{ab}=\frac{\partial\t x^c}{\partial x^k}(\Gamma^k_{ij}
\frac{\partial x^i}{\partial\t x^a}\frac{\partial x^j}{\partial\t x^b}+
\frac{\partial^2 x^k}{\partial\t x^a \partial\t x^b}).
\label{tilde:chris}
\end{equation}

Let us introduce linear operators of covariant differentiation
$\nabla_1,\ldots,\nabla_n$ acting on tensor fields on
$\V$ by the following rules:
\medskip

\noindent 1. If $t=t(x),\ x\in\V$ is a scalar function (invariant), then
$$
\nabla_k t=\partial_k t.
$$

\noindent 2. If $t^i$ is a vector field on $\V$, then
$$
\nabla_k t^i\equiv t^i_{;k}=\partial_k t^i + \Gamma^i_{kj} t^j.
$$

\noindent 3. If $t_i$ is a covector field on $\V$, then
$$
\nabla_k t_i\equiv t_{i;k}=\partial_k t_i - \Gamma^j_{ki} t_j.
$$

\noindent 4. If $u=u^{i_1\ldots i_k}_{j_1\ldots j_l},\
v=v^{i_1\ldots i_r}_{j_1\ldots j_s}$ are tensor fields on $\V$, then
$$
\nabla_k(u\otimes v)=(\nabla_k u)\otimes v + u\otimes\nabla_k v.
$$
\medskip

With the aid of these rules it is easy to calculate covariant
derivatives of arbitrary tensor fields. For example,
for the twice covariant tensor field
$t_{ij}$
$$
\nabla_k t_{ij}\equiv t_{ij;k}=
\partial_k t_{ij} - \Gamma^l_{ik} t_{lj} - \Gamma^l_{jk} t_{il}.
$$
Also, it is easy to check the correctness of the following formulas:
$$
\nabla_k g_{ij}=0,\quad \nabla_k g^{ij}=0,\quad \nabla_k\delta^i_j=0.
$$
Operators of covariant differentiation commute with operations of
symmetrization, alternation, contraction
and also with index rising and index lowing with the aid of metric
tensor.


\subsection{Operators $\Upsilon_k$.}
Let us introduce linear operators
$\Upsilon_1,\ldots,\Upsilon_n$ (Upsilon)
by the following rules:
\medskip

\noindent 1. If $t_{i_1\ldots i_r}$ is a covariant tensor field on
$\V$ of the rank
$r\geq0$, then
$$
\Upsilon_k t_{i_1\ldots i_r}=\partial_k t_{i_1\ldots i_r}.
$$

\noindent 2. $\Upsilon_k dx^i = -\Gamma^i_{kj} dx^j$.
\smallskip

\noindent 3. If $U,V\in\LV$ and $UV$ is a Clifford multiplication of
forms, then
$$
\Upsilon_k(UV)=(\Upsilon_k U)V+ U\Upsilon_k V.
$$
\medskip

With the aid of these rules it is easy to calculate how operators
$\Upsilon_k$ act on arbitrary forms from
$\LV$. For example,
if
$U=\frac{1}{2}u_{kj}dx^k\wedge dx^j=\frac{1}{2}u_{kj}dx^k dx^j$, $u_{kj}=-u_{jk}$,
then
\begin{eqnarray*}
2\Upsilon_i U&=&(\Upsilon_i u_{kj})dx^k dx^j + u_{kj}(\Upsilon_i dx^k)dx^j +
u_{kj} dx^k \Upsilon_i dx^j\\
&=&\partial_i u_{kj}dx^k dx^j+u_{kj}(-\Gamma^k_{il}dx^l)dx^j+
u_{kj}dx^k(-\Gamma^j_{ir}dx^r)\\
&=&(\partial_i u_{kj}-\Gamma^l_{ik}u_{lj}-\Gamma^r_{ij}u_{kr})dx^k\wedge
dx^j.
\end{eqnarray*}

Similarly for
$U\in\Lambda^k(\V)$, written as (\ref{kform}),
we get
\be
\Upsilon_r U=\kfac u_{i_1\ldots i_k;r}dx^{i_1}\wedge\ldots\wedge dx^{i_k}.
\label{UpsilonU}
\ee
If $U\in\LkV$, then
$\Upsilon_r U$ is a covector with the value in
$\Lambda^{k}(\V)$, that is
$\Upsilon_r U=\in\top_1\Lambda^k(\V)$.
The formula (\ref{UpsilonU}) indicate the connection between operators
$\Upsilon_r$ and $\nabla_r$.

Taking into account the third rule in the definition of operators
$\Upsilon_k$, we shall call them {\sl operators of Clifford
differentiation}.

For the change of coordinates $(x)\to(\t x)$
$$
p^i_j=\frac{\partial\t x^i}{\partial x^j},\
q^i_j=\frac{\partial x^i}{\partial\t x^j},\
dx^i=q^i_j d\t x^j,
$$
where $p^i_j,q^i_j$ are functions of  $x\in\V$, the operators of
Clifford differentiation
$\Upsilon_j$ in coordinates $x^1,\ldots,x^n$ related with the operators
of Clifford differentiation $\t\Upsilon_j$ in coordinates
$\t x^1,\ldots,\t x^n$ by the formula
\begin{equation}
\Upsilon_j=p^i_j \t\Upsilon_i
\label{tilde:nabla}
\end{equation}
exact the same as formula for partial derivatives
$\partial_j=p^i_j \t\partial_i$, where $\partial_j=\frac{\partial}{\partial
x^j}$, $\t\partial_j=\frac{\partial}{\partial\t x^j}$.
In order to prove the formula
(\ref{tilde:nabla}) we have to prove the equality
$$
\Upsilon_j dx^k=p^l_j \t\Upsilon_l(q^k_i d\t x^i)
$$
It is easy to check, that the correctness of the last equality follows
from the transformation rule of Christoffel symbols
(\ref{tilde:chris}).

The main properties of the operators $\Upsilon_j$,$j=1,\ldots,n$ are listed
below.

1) $\Upsilon_j I=0$,  where $I=\sqrt{|g|}\,dx^1\ww dx^n$ is the volume
form.

2) $\Upsilon_j(U^*)=(\Upsilon_j U)^*$ for $\forall U\in\LV$.

3) $\Upsilon_j(\star U)=\star(\Upsilon_j U)$ for $\forall U\in\LV$.

4) $\Upsilon_j(\tr\,U)=\tr(\Upsilon_j U)$ for $\forall U\in\LV$.

5) $\Upsilon_j(U,V)=(\Upsilon_j U,V)+(U,\Upsilon_j V)$
 for $\forall U,V\in\LV$.

From the formula $\Upsilon_i dx^k=-\Gamma^k_{ij}dx^j$ we get
\begin{equation}
(\Upsilon_i\Upsilon_j-\Upsilon_j\Upsilon_i)dx^k=R^{\cdot\cdot\cdot k}_{ij,r}dx^r,
\label{R1}
\end{equation}
where
\begin{equation}
R^{\cdot\cdot\cdot k}_{ij,r}=-2(\partial_{[i}\Gamma^k_{j]r}
+\Gamma^k_{s[i}\Gamma^s_{j]r})
\label{R2}
\end{equation}
is (1,3) -- tensor, known as curvature tensor (or Riemannian tensor),
which is equal to zero only in case when the manifold
$\V$ is locally Euclidean.
The curvature tensor with all low indices
$$
R_{ij,rl}=g_{kl} R^{\ldots k}_{ij,r}
$$
is antisymmetric with respect to permutation of indices
$i\lra j$, and also
$r\lra l$. In addition, it symmetric with respect to permutation of
pairs of indices
$ij\lra rl$.
For what follows, it is suitable to introduce tensor from
$\top_2\Lambda^2_\R(\V)$
\begin{equation}
D_{rl}=\frac{1}{2}R_{ij,rl} dx^i\wedge dx^j,
\label{curv:forms}
\end{equation}


\subsection{Operators $d,\delta,\Upsilon,\Delta$.}
With the aid of operators of Clifford differentiation one can define
the operators that map
$\Lambda(\V)$ into $\Lambda(\V)$. We take operators
$d$ and $\Upsilon$ as initial operators. For $V\in\LV$
\begin{eqnarray*}
dV&=&dx^k\wedge\Upsilon_kV,\\
\Upsilon V&=&dx^k \Upsilon_kV.
\end{eqnarray*}
The operator
$\Upsilon$ maps $\Lambda^k(\V) (k=0,\ldots,n)$ into
$\Lambda^{k+1}(\V)\oplus\Lambda^{k-1}(\V)$.

An operator $d$ is called an exterior differential. It has the following
properties:

1) $d\,:\,\Lambda^k(\V)\to\Lambda^{k+1}(\V)$, $k=0,\ldots,n-1$ and
$d\,:\,\Lambda^n(\V)\to 0$.

2) $d^2=0$.

3) $d(U\wedge V)=dU\wedge V+(-1)^k U\wedge dV$, for $U\in\Lambda^k(\V),\,
V\in\Lambda(\V)$

4) If $U\in\Lambda^k(\V)$ is written in a form
$$
U=\kfac u_{i_1\ldots i_k}dx^{i_1}\ww dx^{i_k},\quad
u_{i_1\ldots i_k}=u_{[i_1\ldots i_k]}.
$$
Then
$$
dU=\kfac \partial_{[j}u_{i_1\ldots i_k]}dx^j\wedge dx^{i_1}\ww dx^{i_k}.
$$

Now we may define an operator
$\delta$ by the formula
$$
\delta=d-\Upsilon.
$$
The operator $\delta$ is called a generalized divergence.
It has the following properties:
1) $\delta\,:\,\Lambda^k(\V)\to\Lambda^{k-1}(\V)$,  $k=1,\ldots,n$ and
$\delta\,:\,\Lambda^0(\V)\to 0$.

2) $\delta^2=0$.

3) If $U\in\Lambda^k(\V), (k=1,\ldots,n)$, then
$$
\delta U=(-1)^{k}\star^{-1}d\star U=(-1)^{n(k+1)+1}\sgn(g)\star d\star U.
$$

Finally, let us define an operator
$\Delta$ by the formula
$$
\Delta=\Upsilon^2=(d-\delta)^2=-(d\delta+\delta d).
$$
The operator $\Delta$ is called Beltrami-Laplace operator. It has the
following properties:

1) $\Delta\,:\,\Lambda^k(\V)\to\Lambda^k(\V),\ k=0,\ldots,n$.

2) The operator $\Delta$ commute with operators $d,\delta,\Upsilon,\star$.

3) If $\phi\in\Lambda^0(\V)$, then
$$
\Delta\phi=-\delta
d\phi=\frac{1}{\sqrt{|g|}}\partial_i(\sqrt{|g|}g^{ij}\partial_j\phi).
$$
\medskip

Let's indicate a connection between operators
$d,\delta,\Upsilon$ and the formula (\ref{U1Vk}). For this we may substitute
$\s{1}{U}=dx^j\Upsilon_k$ formally in (\ref{U1Vk}).
As a result, we get a relation
$$
dx^j\Upsilon_j\s{k}{V}=
dx^j\wedge\Upsilon_j\s{k}{V}-(-1)^{k}\star^{-1}(dx^j\wedge\Upsilon_j(\star\s{k}{V})),
$$
which can be written as
$$
\Upsilon\s{k}{V}=d\s{k}{V}-\delta\s{k}{V},
$$
with correspondence with the identity
 $\Upsilon=d-\delta$.

\section{Model equations on Riemannian manifold.}

\subsection{The main equation with two gauge fields and a conservation
law.}
Let $\V$ be and $n$-dimensional elementary Riemannian manifold with
the metric tensor $g_{ij}$.
The main equation of our model is

\begin{equation}
i dx^k(\Upsilon_k\Psi-\Psi a_k-\Psi B_k)-m\Psi=0,
\label{eq:main}
\end{equation}
where $\Psi\in\LCV$; $a_j\in \uone(\V)$
(in other words,  $a_j$ are pure imaginary components of covector);
$B_j\in\spin(\V)$;
$m\in\R$.

We suppose, and in what follows we discuss it, that for $n=4$ equation
(\ref{eq:main}) describes a dynamic of fermion(a spin $1/2$ particle)
of mass $m$ with the presence of electromagnetic field with potential
$a_k$ and gravitational field with potential
$B_k$. A form $\Psi$ is a wave function of fermion. The
equation (\ref{eq:main}) is similar to Dirac equation for electron. A
connection between them is discussed in the chapter 3.5.
For the case of Minkowski space some properties of
the equation (\ref{eq:main}) was
investigated
in \cite{marchuk},\cite{marchuk1}.

The consideration of properties of the equation
(\ref{eq:main}) in $\V$ we begin with a conservation law. Two lemmas will be
used.

\lemma 1. If $C\in\LCV,\ H\in\Lambda^1_\R(\V)$, then the scalar
functions of
$x\in\V$
$$
\tr(H(C+C^*)),\ \tr(iH(C-C^*))
$$
are real valued.\par

\proof. Let $C=\s{0}{C}+\cdots+\s{n}{C},\ \s{k}{C}\in\Lambda^k_\C(\V)$.
Substituting $\s{1}{U}=H,\ \s{k}{V}=\s{k}{C}$ into the formula (\ref{U1Vk})
we get
$$
H\s{k}{C}\in\Lambda^{k+1}_\C(\V)\oplus\Lambda^{k-1}_\C(\V).
$$
Therefore $\tr(HC)=\tr(H\s{1}{C})$. We may write
$\s{1}{C}=(r_k+is_k)dx^k$,  $r_k,s_k\in\R$. Then
\begin{eqnarray*}
\tr(H(\s{1}{C}+\s{1}{C^*}))&=&\tr(H 2r_k dx^k)\in\R,\\
\tr(iH(\s{1}{C}-\s{1}{C^*}))&=&\tr(-H 2s_k dx^k)\in\R\fin
\end{eqnarray*}
\medskip

Let us denote
\begin{equation}
C=\Psi^*(idx^k(\Upsilon_k\Psi-\Psi a_k-\Psi B_k)-m\Psi)
\label{C}
\end{equation}
and apply the operation of conjugation
\begin{equation}
C^*=(-(\Upsilon_k\Psi^*+a_k\Psi^*+B_k\Psi^*)idx^k-m\Psi^*)\Psi.
\label{C*}
\end{equation}
Here we have taken into account, that $a_k^*=-a_k$, $B_k^*=-B_k$,
$(\Upsilon_k\Psi)^*=\Upsilon_k\Psi^*$. The important role in our
construction play the real 1-form
$H\in\Lambda^1_\R(\V)$, satisfying the equations
\begin{equation}
H^2=1,\quad\Upsilon_k H=HB_k-B_k H,\quad k=1,\ldots,n
\label{eq:H}
\end{equation}
Later on a consideration of conditions  of solvability of this system
of equations leads us to important implications.

\lemma 2. If $C^*$ is defined by the formula (\ref{C*}), and
$H\in\Lambda^1_\R(\V)$ satisfies equations (\ref{eq:H}), then the form
$\bar{C}\equiv HC^*$ can be written as
$$
\bar{C}=(-(\Upsilon_k\bar\Psi+a_k\bar\Psi+B_k\bar\Psi)idx^k-m\bar\Psi)\Psi,
$$
where $\bar\Psi\equiv H\Psi^*$.
\par

\proof\,\, is follows from the equalities
\begin{eqnarray*}
HC^*&=&(-(H\Upsilon_k\Psi^*+a_k H\Psi^*+HB_k\Psi^*)idx^k-mH\Psi^*)\Psi\\
&=&(-(\Upsilon_k\bar\Psi+a_k \bar\Psi+B_k\bar\Psi+
(HB_k-B_k H-\Upsilon_k H)\Psi^*)idx^k-m\bar\Psi)\Psi\fin
\end{eqnarray*}

\theorem 10. Let forms $\Psi\in\LCV$; $a_k\in\uone(\V)$;
$B_k\in\spin(\V)$ satisfy the equation (\ref{eq:main}).
We may consider a vector with components
$$
j^k=\tr(\bar\Psi dx^k\Psi),
$$
where $\bar\Psi=H\Psi^*$, and $H\in\Lambda^1_\R(\V)$ satisfies the
system of equations (\ref{eq:H}). Then, the following identity is true
$$
\partial_k(\sqrt{|g|}j^k)=0,
$$
which is called a conservation law for the equation  (\ref{eq:main}).
The vector with components
$j^1,\ldots,j^n$ is called a current.
\par

\proof. Let a form $C$ is defined by the formula (\ref{C}). Using the
lemma 2, we may write
\begin{eqnarray*}
-iH(C-C^*)&=&\bar\Psi(dx^k(\Upsilon_k\Psi-\Psi a_k-\Psi B_k)+im\Psi)\\
&&+((\Upsilon_k\bar\Psi+a_k\bar\Psi+B_k\bar\Psi)dx^k-im\bar\Psi)\Psi\\
&=&\bar\Psi dx^k\Upsilon_k\Psi+\Upsilon_k\bar\Psi dx^k\Psi-\bar\Psi dx^k\Psi B_k
+B_k\bar\Psi dx^k\Psi.
\end{eqnarray*}
As $\tr(UV-VU)=0$, then
\begin{eqnarray*}
\tr(-iH(C-C^*))&=&\tr(\bar\Psi dx^k\Upsilon_k\Psi+\Upsilon_k\bar\Psi dx^k\Psi)\\
&=&\tr(\Upsilon_k(\bar\Psi dx^k\Psi)-\bar\Psi(\Upsilon_k dx^k)\Psi)\\
&=&\partial_k \tr(\bar\Psi dx^k\Psi)+\tr(\Gamma^k_{kl}\bar\Psi dx^l\Psi)\\
&=&\partial_k j^k+\Gamma^k_{kl}j^l.
\end{eqnarray*}
Using the well known formula
$$
\Gamma^k_{kl}=\partial_l(\ln\sqrt{|g|}),
$$
we get
$$
\partial_k
j^k+\Gamma^k_{kl}j^l=\frac{\partial_k(\sqrt{|g|}\,j^k)}{\sqrt{|g|}}.
$$
Hence, if equation (\ref{eq:main}) is satisfied, then
$C=0$ and we get
$$
\partial_k(\sqrt{|g|}\,j^k)=0.
$$
By the lemma 1,  components of the vector $j^k$ are real valued\fin
\medskip

\remark. If we take 1-form $J=j_k dx^k$, where $j_k=g_{kl}j^l$,
then the equality $\partial_k(\sqrt{|g|}j^k)=0$ can be written using the
operator of generalized divergence
$$
\delta J=0.
$$

\subsection{The gauge invariance and Lagrangian.}
\lemma \cite{benn}. If $B_j\in\spin(\V)$, and¨ $U\in\Spin(\V)$. Then
$U^{-1}B_j U\in\spin(\V)$ and
$U^{-1}\Upsilon_j U\in\spin(\V)$\fin
\par

\theorem 11. The equation (\ref{eq:main}) is invariant over the
following gauge transformation:
\begin{eqnarray*}
\Psi&\to&\Psi^\prime=\Psi Uv,\\
a_j&\to&a_j^\prime=a_j+v^{-1}\partial_j v,\\
B_j&\to&B_j^\prime=U^{-1}B_j U+U^{-1}\Upsilon_j U,
\end{eqnarray*}
where $v\in\UV$, $U\in\Spin(\V)$ and $Uv=vU$.
\par

\proof. Substituting to the left part of equation (\ref{eq:main})
forms $\Psi^\prime,a_j^\prime,B_j^\prime$ instead of $\Psi,a_j,B_j$, we
get
$$
i dx^j(\Upsilon_j\Psi^\prime-\Psi^\prime a_j^\prime-\Psi^\prime B_j^\prime)-
m\Psi^\prime=
(i dx^j(\Upsilon_j\Psi-\Psi a_j-\Psi B_j)-m\Psi)Uv=0\fin
$$

In Quantum Field Theory it is important, that the main equations (Dirac,
Klein-Gordon, Maxwell, Yang-Mills etc.) are derived from the respective
Lagrangians with the aid of variational principle
\cite{bogolubov}. In accordance with it, the next step of our program is
to specify the Lagrangian from which the equation
(\ref{eq:main}) can be derived. For this, we shall use a real 1-form
$H\in\Lambda^1_\R(\V)$, which satisfies the system of equation
(\ref{eq:H}).

\theorem 12. The equation (\ref{eq:H}) are invariant over the gauge
transformation
\begin{eqnarray*}
H&\to&H^\prime=U^{-1}HU,\\
B_j&\to&B_j^\prime=U^{-1}B_j U+U^{-1}\Upsilon_j U,
\end{eqnarray*}
where $U\in\Spin(\V)$.
\par

\proof. We may multiply the left parts of equations
$\Upsilon_j H-HB_j+B_j H=0$ from left on
$U^{-1}$  and from right on $U$. Then, using the equality
$(\Upsilon_j U^{-1})U=-U^{-1}\Upsilon_j U$, we get
$$
U^{-1}(\Upsilon_j H-HB_j+B_j H)U=\Upsilon_j H^\prime-H^\prime B_j^\prime+
B_j^\prime H^\prime\fin
$$

Now we can write down the Lagrangian of interest
\begin{eqnarray}
\L_1&=&\L_1(\Psi,\bar\Psi,a_j,B_j)=\tr(H(C+C^*))\nonumber\\
&=&\tr\{\bar\Psi(idx^k(\Upsilon_k\Psi-\Psi a_k-\Psi B_k)-m\Psi)+\label{lagr1}\\
&&(-(\Upsilon_k\bar\Psi+a_k\bar\Psi+B_k\bar\Psi)idx^k-m\bar\Psi)\Psi\}.\nonumber
\end{eqnarray}
where $C,C^*$ are defined by (\ref{C}),(\ref{C*}).

\theorem 13. The Lagrangian (\ref{lagr1}) is invariant over the
following gauge transformation:
\begin{eqnarray}
\Psi&\to&\Psi^\prime=\Psi Uv,\nonumber \\
\bar\Psi&\to&\bar\Psi^\prime=v^{-1}U^{-1}\bar\Psi, \label{gauge}\\
a_j&\to&a_j^\prime=a_j+v^{-1}\partial_j v,\nonumber\\
B_j&\to&B_j^\prime=U^{-1}B_j U+U^{-1}\Upsilon_j U,\nonumber
\end{eqnarray}
where $v\in \UV$, $U\in\Spin(\V)$.
\par

\proof. We can write the Lagrangian (\ref{lagr1}) as
$$
\L_1(\Psi,\bar\Psi,a_j,B_j)=\tr(Q(\Psi,\bar\Psi,a_j,B_j)),
$$
where $Q$ is the corresponding form from (\ref{lagr1}).
Using the theorem 11, it is easy to check, that
$$
\tr(Q(\Psi^\prime,\bar\Psi^\prime,a_j^\prime,B_j^\prime))
=\tr(U^{-1}Q(\Psi,\bar\Psi,a_j,B_j)U)
=\tr(Q(\Psi,\bar\Psi,a_j,B_j))\fin
$$
\medskip

Now we may complete  the Lagrangian (\ref{lagr1}) by terms, which
describe free fields
$a_j\in \uone(\V)$ and
$B_j\in\spin(\V)$. For this, we are taking tensors
$f_{ij}\in \uonetwo(\V)$ and
$G_{ij}\in\spintwo(\V)$
\begin{eqnarray}
f_{ij}&=&\Upsilon_i a_j-\Upsilon_j a_i = \partial_i a_j-\partial_j
a_i,\nonumber\\
G_{ij}&=&\Upsilon_i B_j-\Upsilon_j B_i+B_i B_j-B_j B_i
\label{fG}
\end{eqnarray}
to compose the Lagrangian
\begin{equation}
\L_0=\L_0(a_j,B_j)=\tr(c_1\sqrt{|g|}f_{ij}f^{ij}+c_2\sqrt{|g|}G_{ij}G^{ij}),
\label{lagr0}
\end{equation}
where $f^{ij}=g^{ir}g^{js}f_{rs}$, $G^{ij}=g^{ir}g^{js}G_{rs}$,
$c_1,c_2$--real constants. It is evident, that Lagrangian
(\ref{lagr0})  is also invariant over the gauge transformation
(\ref{gauge}) with
$$
f_{ij}\to f_{ij}^\prime=f_{ij},\quad
G_{ij}\to G_{ij}^\prime=U^{-1}G_{ij}U.
$$
The complete Lagrangian $\L$  is a sum of Lagrangians $\L_0$ and $\L_1$
\begin{eqnarray}
\L&=&\L(\Psi,\bar\Psi,a_j,B_j)=\L_0+\L_1 \label{lagr}\\
&=&\tr\{\bar\Psi(idx^j(\Upsilon_j\Psi-\Psi a_j-\Psi B_j)-m\Psi)\nonumber\\
&&+(-(\Upsilon_j\bar\Psi+a_j\bar\Psi+B_j\bar\Psi)idx^j-m\bar\Psi)\Psi\nonumber\\
&&+c_1\sqrt{|g|}f_{ij}f^{ij}+c_2\sqrt{|g|}G_{ij}G^{ij}\},\nonumber
\end{eqnarray}
where $f_{ij},G_{ij}$ are defined by formulas (\ref{fG}). The
Lagrangian $\L$ is invariant over the gauge transformation
(\ref{gauge}) with gauge groups $\UV$ and $\Spin(\V)$.
Let us write the forms $\Psi,\bar\Psi,a_j,B_j$ as decompositions over
the basis
(\ref{dx:basis}), considering forms $\Psi$ and $\bar\Psi$ as independent
forms. Then Lagrangian
$\L=\L(\Psi,\bar\Psi,a_j,B_j)$ is a scalar valued and depends on the
coefficients of forms
$\Psi,\bar\Psi,a_j,B_j$
and on their partial derivatives. The Lagrangian $\L$,
generally speaking, is complex valued,  and for variation we must take a
real part
$\L_\R={\rm Re}\,\L$ (by lemma 1, $\L$ is real if
$\bar\Psi=H\Psi^*$, where $H$ is a solution of equation (\ref{eq:H})).
To derive a system of equations from the Lagrangian
$\L_\R$, we have to write down the Lagrange-Euler equations
$$
\frac{\partial\L_\R}{\partial
u_\alpha}-\partial_k\frac{\partial\L_\R}{\partial u_{\alpha;k}}=0,
$$
where $u_{\alpha;k}=\partial_k u_\alpha$, and¨ $u_\alpha=u_\alpha(x)$
all real and imaginary parts of the coefficients of forms
$\bar\Psi,a_j,B_j$,$(j=1,\ldots,n)$.
It can be checked (we have done it for $n\leq4$), that the resulting
system of equations can be written again with the aid of forms
$\Psi,\bar\Psi,a_j,B_j$:
\begin{eqnarray}
&&idx^j(\Upsilon_j\Psi-\Psi a_j-\Psi B_j)-m\Psi=0,\nonumber\\
&&\frac{1}{\sqrt{|g|}}\partial_i(\sqrt{|g|}f^{ij})=
\frac{1}{c_1}(J^j)_{\uoneup(\V)},\label{main}\\
&&\frac{1}{\sqrt{|g|}}\Upsilon_i(\sqrt{|g|}G^{ij})-[G^{ij},B_i]=
\frac{1}{c_2}(J^j)_{\spinup(\V)},\nonumber
\end{eqnarray}
where $f_{ij},G_{ij}$ are defined in (\ref{fG}),
 $J^j=i\bar\Psi dx^j\Psi$, and $(J^j)_{\uoneup(\V)},\
 (J^j)_{\spinup(\V)}$ are projections of  the values
$J^j$ on $\uoneup(\V)$ and
$\spinup(\V)$.
In other words, if we write forms $J^j$ as decompositions over the basis
(\ref{dx:basis}),
$$
J^j=p^j+p^j_i dx^i + \sum_{i_1<i_2}p^j_{i_1 i_2}dx^{i_1}\wedge dx^{i_2}+\cdots,
$$
then
$$
(J^j)_{\uoneup(\V)}=i{\rm Im}\,p^j,\quad
(J^j)_{\spinup(\V)}=
\sum_{i_1<i_2}{\rm Re}(p^j_{i_1 i_2})dx^{i_1}\wedge dx^{i_2}.
$$
Finally, assuming that in the system of equations
(\ref{main}) $\bar\Psi=H\Psi^*$ we may complete it by the equations
(\ref{eq:H}) for
$H\in\Lambda^1_\R(\V)$.


\subsection{Conditions of a solvability of equations for $H$ and a general
system of equations.}
Let us consider in more details equations (\ref{eq:H}) and conditions of
their solvability. If the form
$H=h_i dx^i\in\Lambda^1_\R(\V)$ has twice differentiable coefficients
$h_i=h_i(x),\ x\in\V$, then the following relation is true:
$$
\Upsilon_i(\Upsilon_j H)-\Upsilon_j(\Upsilon_i H)=\Upsilon_i(HB_j-B_j H)
-\Upsilon_j(HB_i-B_iH),\ i,j=1,\ldots,n.
$$
It can be rewritten in a form
\begin{equation}
(\Upsilon_i\Upsilon_j-\Upsilon_j\Upsilon_i)H=HG_{ij}-G_{ij}H,\quad i,j=1,\ldots,n
\label{HG}
\end{equation}
where
$$
G_{ij}=\Upsilon_i B_j-\Upsilon_j B_i+B_i B_j-B_j B_i
$$
are tensors from $\spintwo(\V)$
$$
G_{ij}=\frac{1}{2}b_{kl,ij}dx^k\wedge dx^l,\quad
b_{kl,ij}=b_{[kl],ij}=b_{kl,[ij]}.
$$

\theorem 14. If the relations (\ref{HG}) is true for all
$H\in\Lambda^1_\R(\V)$, then
\begin{equation}
b_{ij,kl}=-\frac{1}{2}R_{ij,kl},\quad i,j=1,\ldots,n
\label{bR}
\end{equation}
where $R_{ij,kl}=g_{ls}R^{\cdot\cdot\cdot s}_{ij,k}$
are components of a curvature tensor
with low indices.
\par

\proof. Let us denote $e^k=dx^k,\ k=1,\ldots,n$. By definition of the
curvature tensor (\ref{R1})
$$
(\Upsilon_i\Upsilon_j-\Upsilon_j\Upsilon_i)e^k=R^{\cdot\cdot\cdot k}_{ij,l}e^l.
$$
Substituting $H=e^k$ into the equality (\ref{HG}), we get
\begin{eqnarray*}
R^{\cdot\cdot\cdot k}_{ij,l}e^l&=&e^k G_{ij}-G_{ij}e^k=
\frac{1}{2}b_{ij,rs}(e^k e^r e^s-e^r e^s e^k)\\
&=&\frac{1}{2}b_{ij,rs}((2g^{kr}-e^r e^k)e^s-e^r(2g^{ks}-e^k e^s))=
b_{ij,rs}(g^{kr}e^s-g^{ks}e^r)\\
&=&g^{kr}b_{ij,rs}e^s-g^{ks}b_{ij,rs}e^r=
2g^{kr}b_{ij,rs}e^s=-2b^{\cdot\cdot\cdot k}_{ij,s}e^s.
\end{eqnarray*}
Comparing the begin and the end of this chain of equalities,
we get (\ref{bR})\fin
\medskip

The formula (\ref{bR}) can be written as a relation between tensors $G_{ij}$
and forms (\ref{curv:forms})
\begin{equation}
G_{ij}=-\frac{1}{2}D_{ij},\quad i,j=1,\ldots,n.
\label{GC}
\end{equation}
We include this relation into the system of
equations (\ref{main}),(\ref{eq:H}) for the description of
influence of
fields $\Psi,a_j,B_j$ to the curvature of Riemannian manifold $\V$.
\medskip

\noindent {\bf Postulate. \sl We suppose, that the curvature tensor of
Riemannian manifold
$\V$ is connected with the solution of system of
equations (\ref{main}),(\ref{eq:H}) by the formula
(\ref{GC})}.
\medskip

This connection gives us possibility to interpret the
tensor $B_j$ as a potential of gravitational field, and the tensor
$G_{ij}$ as a strength of gravitational field.
\medskip

Now, we may write down a general system of equations, which connect the
following three fields: a field of matter, described by a wave function
of fermion $\Psi\in\LCV$; an electromagnetic field, described by a
potential
$a_j\in \uone(\V)$  and by strength
$f_{ij}\in  \uonetwo(\V)$; a gravitational field, described by a potential
$B_j\in\spin(\V)$ and by strength
$G_{ij}\in\spintwo(\V)$:
\begin{eqnarray}
&&idx^j(\Upsilon_j\Psi-\Psi a_j-\Psi B_j)-m\Psi=0,\nonumber\\
&&f_{ij}=\partial_i a_j-\partial_j a_i,\quad i,j=1,\ldots,n\nonumber\\
&&G_{ij}=\Upsilon_i B_j-\Upsilon_j B_i+B_i B_j-B_j B_i,\quad
i,j=1,\ldots,n\label{main1}\\
&&\frac{1}{\sqrt{|g|}}\partial_i(\sqrt{|g|}f^{ij})=
\frac{1}{c_1}(J^j)_{\uoneup(\V)},\quad j=1,\ldots,n\nonumber\\
&&\frac{1}{\sqrt{|g|}}\Upsilon_i(\sqrt{|g|}G^{ij})-[G^{ij},B_i]=
\frac{1}{c_2}(J^j)_{\spinup(\V)},\quad j=1,\ldots,n\nonumber\\
&&\bar\Psi=H\Psi^*,\quad J^j=i\bar\Psi dx^j\Psi,\quad j=1,\ldots,n\nonumber\\
&&H^2=1,\quad \Upsilon_j H = H B_j-B_j H,\quad j=1,\ldots,n\nonumber\\
&&D_{ij}=\frac{1}{2}R_{rl,ij}dx^r\wedge dx^l=-2G_{ij},\quad i,j=1,\ldots,n\nonumber
\end{eqnarray}
where $H\in\Lambda^1_\R(\V)$, $R_{rl,ij}$ is a curvature tensor with
low indices and $U\in\Spin(\V)$.


\subsection{A covariance of main system of  equations under a
Spin-isometric change of coordinates.}
Let us prove a covariance of equation
\begin{equation}
i dx^j(\Upsilon_j\Psi-\Psi a_j-\Psi B_j)-m\Psi=0,
\label{cov1}
\end{equation}
under the change of coordinates $(x)\to(\t x)$ on $\V$
$$
x^j=x^j(\t x^1,\ldots,\t x^n),\quad \t x^j=\t x^j(x^1,\ldots,x^n)
\quad j=1,\ldots,n
$$
\be
\frac{\partial\t x^i}{\partial x^j}=p^i_j,\quad
\frac{\partial x^i}{\partial\t x^j}=q^i_j,\quad
p^i_j q^j_k=\delta^i_k,\quad
p^i_j q^k_i=\delta^k_j
\label{change:coord}
\ee
where $\delta^i_j$ is a Kronecker tensor, $\delta^i_i=1$ and $\delta^i_j=0$
for $i\neq j$. Substituting $dx^j=q^j_k d\t x^k$,
$\Upsilon_j=p^l_j\t\Upsilon_l$ into the equation (\ref{cov1}), and using the
relations
$$
dx^j\Upsilon_j=d\t x^k q^j_k p^l_j \t\Upsilon_l=d\t x^k\t\Upsilon_k,
$$
we get
\begin{equation}
i d\t x^k(\t\Upsilon_k\Psi-\Psi\t a_k-\Psi\t B_k)-m\Psi=0,
\label{cov2}
\end{equation}
where
\begin{equation}
\t a_k=a_j q^j_k,\quad \t B_k=B_j q^j_k.
\label{cov3}
\end{equation}
because $a_j\in\uone(\V)$, $B_j\in\spin(\V)$.

A form $\Psi\in\LCV$ is a nonhomogeneous covariant antisymmetric tensor field
$$
\Psi=\sum_{k=0}^n\kfac \psi_{i_1\ldots i_k}dx^{i_1}\ww dx^{i_k},\quad
\psi_{i_1\ldots i_k}=\psi_{[i_1\ldots i_k]}
$$
Therefore, in new coordinates the form $\Psi$ writes
\begin{equation}
\t\Psi=
\sum_{k=0}^n\kfac \psi_{i_1\ldots i_k}q^{i_1}_{j_1}\ldots q^{i_k}_{j_k}
d\t x^{j_1}\ww d\t x^{j_k},
\label{tilde:psi}
\end{equation}
and the  equation (\ref{cov1}) becomes
\begin{equation}
i d\t x^k(\t\Upsilon_k\t\Psi-\t\Psi\t a_k-\t\Psi\t B_k)-m\t\Psi=0,
\label{cov4}
\end{equation}
that proves the covariance of equation
(\ref{cov1}). Now, let us consider a Spin-isometric change of
coordinates (\ref{change:coord}
associated with some form $F\in\Spin(1,3)$
$$
dx^j=q^j_k d\t x^k=Fd\t x^j F^{-1}.
$$
By the theorem 8, the form $\Psi$ in coordinates $(\t x)$ has a
following view:
$$
\t\Psi=F(\sum_{k=0}^n\kfac \psi_{i_1\ldots i_k}
d\t x^{i_1}\ww d\t x^{i_k})F^{-1}=F\breve\Psi F^{-1}.
$$
The equation (\ref{cov4}) can be written as
\begin{equation}
i d\t x^k(\t\Upsilon_k(F\breve\Psi F^{-1})-
(F\breve\Psi F^{-1})\t a_k-(F\breve\Psi F^{-1})\t B_k)-
m(F\breve\Psi F^{-1})=0,
\label{cov5}
\end{equation}
For this equation we can do a gauge transformation with the same
form $F\in\Spin(\V)$
\begin{eqnarray*}
\t\Psi&\to&\t\Psi^\prime=\t\Psi F=F\breve\Psi,\\
\t B_k&\to&\t B_k^\prime=F^{-1}\t B_k F+F^{-1}\t\Upsilon_k F,\\
\t a_k&\to&\t a_k^\prime=\t a_k.
\end{eqnarray*}
We get
\begin{equation}
i d\t x^k(\t\Upsilon_k(F\breve\Psi)-
(F\breve\Psi)\t a_k^\prime-(F\breve\Psi)\t B_k^\prime)-
m(F\breve\Psi)=0,
\label{cov6}
\end{equation}

Hence, if we do successively a Si-change of coordinates associated with
$F\in\Spin(\V)$, and the gauge transformation with the same
$F$, then we get the equation (\ref{cov6})
that looks like (\ref{cov1}), but with the different
transformation rule for the wave
function
\begin{equation}
\Psi\to F\breve\Psi.
\label{cov7}
\end{equation}
which is characteristic for spinors, but not for tensors.
\medskip

This consideration leads us to the conclusion, that in our model we must
accept only Spin-isometric change of coordinates on $\V$. And so, we
come to a new variety of geometry of Riemannian manifold.

\df. An elementary Riemannian manifold $\V$ is called Spin-isometric
manifold (Si-manifold), if we can do only Si-changes of coordinates on
it.\par

We may introduce the suggestion about the structure of physical
space.
\medskip

\noindent {\bf Postulate. \sl The physical space locally (at the
neighborhood of every fixed point) is a four dimensional Spin-isometric
manifold with a metric tensor of the signature $-2$, or $+2$.}
\medskip


\subsection{The relation of main equation of our model
in Minkowski space with
Dirac equation.}
Let $\E$ be a Minkowski space ($n=4$) with coordinates
$x=(x^1,x^2,x^3,x^4)$, where $x^4$ is time coordinate, with the basis
coordinate vectors
$e_1,e_2,e_3,e_4$ and the basis covectors
$e^i=g^{ij}e_j$, where
$$
\|g^{ij}\|=\|g_{ij}\|=\diag(-1,-1,-1,1).
$$
Dirac equation for an electron has a form
\begin{equation}
(i\gamma^k(\partial_k-a_k)-m)\theta=0,
\label{eq:dirac}
\end{equation}
where $m\in\R$; $\gamma^1,\ldots,\gamma^4$ are complex valued Dirac
matrices of the dimension $4\!\times\!4$, which satisfy conditions
$\gamma^k\gamma^l+\gamma^l\gamma^k=2g^{kl}{\bf1}$ ($\bf1$ is identity
matrix), that means $\gamma^1,\ldots,\gamma^4$ are matrix representation
of generators $e^1,\ldots,e^4$ of Clifford algebra $\cl_\C(1,3)$;
$\theta=\theta(x)$ is a column of four complex valued functions (Dirac
spinor); the pure imaginary functions
$a_j=a_j(x)$ describe a potential of electromagnetic field. For
$\gamma^k$ we may take a concrete matrix representation, for example,
Dirac's representation
$$
\gamma^1=\pmatrix{0 &0 &0 &-1\cr
                  0 &0 &-1&0 \cr
                  0 &1 &0 &0 \cr
                  1 &0 &0 &0},\quad
\gamma^2=\pmatrix{0 &0 &0 & i\cr
                  0 &0 &-i&0 \cr
                  0 &-i&0 &0 \cr
                  i &0 &0 &0},
$$
$$
\gamma^3=\pmatrix{0 &0 &-1& 0\cr
                  0 &0 & 0&1 \cr
                  1 &0 &0 &0 \cr
                  0 &-1&0 &0},\quad
\gamma^4=\pmatrix{1 &0 &0 & 0\cr
                  0 &1 & 0&0 \cr
                  0 &0 &-1&0 \cr
                  0 &0 &0 &-1}.
$$
Here we shall prove a covariance of Dirac equation (\ref{eq:dirac})
under change of coordinates  from
the group $SO^+(1,3)$ (details see, for example, in
\cite{bogolubov})
$$
x^k=q^k_j\tilde x^j,\quad\partial_k=p^j_k\tilde\partial_j,\quad
p^i_j q^j_k=q^i_j p^j_k=\delta^i_k.
$$
As was shown in the chapter 1.13, this change of coordinates is
Spin-isometric, associated with some form
$F\in\Spin(1,3)$
and
\begin{equation}
p^j_k e^k=F^*e^j F,\quad q^k_j e^j=F e^k F^*.
\label{spin:formula}
\end{equation}
If we write down the form $F$ in Clifford basis (\ref{cl:basis})
$$
F=f e+\sum_{i<j}f_{ij}e^{ij}+f_{1234}e^{1234}
$$
then we may take the corresponding matrix
$$
{\bf F}=f{\bf1}+\sum_{i<j}f_{ij}\gamma^i\gamma^j+
f_{1234}\gamma^1\gamma^2\gamma^3\gamma^4.
$$
and rewrite relations (\ref{spin:formula}) as the matrix relations
\begin{equation}
p^j_k \gamma^k={\bf F}^*\gamma^j {\bf F},\quad q^k_j \gamma^j=
{\bf F} \gamma^k {\bf F}^*.
\label{spin2:formula}
\end{equation}
Substituting the equality $\partial_k=p^j_k\tilde\partial_j$ in
(\ref{eq:dirac}) and using the formulas (\ref{spin2:formula}), we get
\begin{eqnarray*}
(i\gamma^k(\partial_k-a_k)-m)\theta
&=&(i\gamma^k(p^j_k\tilde\partial_j-a_k)-m)\theta\\
&=&(i{\bf F}^{-1}\gamma^j {\bf F}\tilde\partial_j-i\gamma^k a_k-m)\theta\\
&=&{\bf F}^{-1}(i\gamma^j\tilde\partial_j-i{\bf F}\gamma^k {\bf
F}^{-1}a_k-m)({\bf F}\theta)\\
&=&{\bf F}^{-1}(i\gamma^j(\tilde\partial_j-\tilde a_j)-m)({\bf F}\theta),
\end{eqnarray*}
where $\tilde a_j=q^k_j a_k$. Hence, the equation (\ref{eq:dirac})
under Lorentzian transformation from $SO^+(1,3)$ gives the equation
\begin{equation}
(i\gamma^j(\tilde\partial_j-\tilde a_j)-m)\tilde\theta=0,
\label{eq:dirac1}
\end{equation}
where $\tilde\theta={\bf F}\theta$ and ${\bf F}$ is a matrix
representation of the element
$F$ of the group $\Spin(1,3)$.

Now we want to compare in Minkowski space the equation
(\ref{eq:dirac}) with the main equation (\ref{eq:main}) in which
$B_k=0$, $\Upsilon_k=\partial_k$ and $dx^k=e^k$
\begin{equation}
(ie^k(\partial_k-a_k)-m)\Psi=0,
\label{kappa}
\end{equation}
where $\Psi\in\Lambda_\C(\E)$, $m\in\R$, $a_1,\ldots,a_4$ are the same
as in (\ref{eq:dirac}). In accordance with the definition of Clifford
multiplication, we have
$e^k e^l+ e^l e^k=2 g^{kl}e$. Therefore, for $e^k$ we may use the matrix
representation $\gamma^k$. If we write the form $\Psi$ as decomposition
(theorem 1)
$$
\Psi=\sum_{k=0}^4\kfac \psi_{i_1\ldots i_k}e^{i_1}\ldots e^{i_k},
\quad \psi_{i_1\ldots i_k}=\psi_{[i_1\ldots i_k]}
$$
then we can associate with it the following matrix
$$
{\bf \Psi}=\sum_{k=0}^4\kfac \psi_{i_1\ldots i_k}\gamma^{i_1}\ldots\gamma^{i_k},
$$
As a result, we get the matrix equation
\begin{equation}
(i\gamma^k(\partial_k-a_k)-m){\bf\Psi}=0,
\label{kappa1}
\end{equation}
which differ from (\ref{eq:dirac}) only by the fact, that
${\bf\Psi}$ is $4\!\times\!4$-matrix, but not a four components column.
One can establish a correspondence between a column
$\theta$ and subclass of
$4\!\times\!4$-matrices of the form
$$
\theta=\pmatrix{\theta_1\cr\theta_2\cr\theta_3\cr\theta_4}
\leftrightarrow\pmatrix{\theta_1&0&0&0\cr
\theta_2&0&0&0\cr\theta_3&0&0&0\cr\theta_4&0&0&0}={\bf\Psi}_\theta
$$
which gives the equivalence of the equation (\ref{eq:dirac}) and the
equation
\begin{equation}
i\gamma^k(\partial_k-a_k)-m){\bf\Psi}_\theta=0.
\label{kappa2}
\end{equation}
Now, let us see how the equation  (\ref{kappa1}) and, in particular,
(\ref{kappa2}) transforms under a change of coordinates. As was shown in
the previous chapter, the equation
(\ref{kappa1}) under the Si-change of coordinates (from the group
$SO^+(1,3)$), associated with a form
$F\in\Spin(1,3)$, transforms into the equation
$$
(i\tilde e^k(\tilde\partial_k-\tilde a_k)-m)(F\breve\Psi)F^{-1}=0,
$$
where
$$
\breve\Psi=
\sum_{k=0}^4 \psi_{i_1\ldots i_k}\tilde e^{i_1}\ldots\tilde e^{i_k}.
$$
Basis covectors
$\tilde e^k$ in coordinates $(\tilde x)$
satisfy the same relations
$\tilde e^k\tilde e^l+\tilde e^l\tilde e^k=2 g^{kl}e$,
and hence, we can associate with them the same matrices $\gamma^k$.
As a result, after multiplication from right on
$F$, we get the equation
$$
(i\gamma^k(\tilde\partial_k-\tilde a_k)-m)({\bf F\Psi})=0,
$$
which precisely corresponds to the transformed equation
(\ref{eq:dirac1}).


\end{document}